
\magnification=1097
\input /u/atd2h/TeX/jnl
\input /u/atd2h/TeX/reforder
\input /u/atd2h/TeX/eqnorder

\def\uva{Department of Physics\\University of Virginia\\
         McCormick Road\\Charlottesville, Virginia 22901}
\def\pacs{\vskip 3pt plus 0.2fill \noindent  PACS numbers: }
\def \del2{{\nabla^{2}}}

\def\r{{\bf r}}
\def\d{{\bf d}}
\def\E{{\bf E}}
\def\B{{\bf B}}
\def\H{{\bf H}}
\def\h{{\bf h}}
\def\A{{\bf A}}
\def\Q{{\bf Q}}
\def\J{{\bf J}}
\def\M{{\bf M}}

\def\Jn{{\bf J}_{n}}
\def\Js{{\bf J}_{s}}
\def\vl{{\bf v}_{L}}
\def\rl{{\bf r}_{L}}
\def\vs{{\bf v}_{s}}
\def\vsi{{\bf v}_{s1}}
\def\f{{\bf f}}
\def\F{{\bf F}}
\def\sigxx{{\sigma_{xx}^{(n)}}}
\def\sigxy{{\sigma_{xy}^{(n)}}}
\def\sigyx{{\sigma_{yx}^{(n)}}}
\def\sig{{\sigma^{(n)}}}
\def\etheta{{\bf e}_{\theta}}
\def\er{{\bf e}_{r}}
\def\ez{{\bf e}_{z}}

\def\gr{{\gamma_{1}}}
\def\gi{{\gamma_{2}}}
\def\eptil{{\tilde{\epsilon}}}


\title Vortex motion and the Hall effect in type II superconductors:
       a time dependent Ginzburg-Landau theory approach

\author Alan T. Dorsey

\affil \uva

\abstract Vortex motion in type II superconductors is studied  starting from
a variant of the time dependent Ginzburg-Landau equations, in which the
order parameter relaxation time is taken to be complex.  Using a method
due to Gor'kov and Kopnin, we derive an equation of motion for a single
vortex ($B\ll H_{c2}$) in the presence of an applied transport current.
The imaginary part of the relaxation time and the normal state
Hall effect both break ``particle-hole
symmetry,'' and produce a component of the vortex velocity parallel to the
transport current,  and consequently a Hall field due to the vortex motion.
Various models for the relaxation time are considered, allowing for a
comparison to some phenomenological models of vortex motion
in superconductors, such as the Bardeen-Stephen and Nozi\`eres-Vinen models,
as well as to models of vortex motion in neutral superfluids.
In addition, the transport energy,
Nernst effect, and thermopower are calculated for a single vortex.
Vortex bending and fluctuations can also be included within this description,
resulting in a Langevin equation description of the vortex motion.
The Langevin equation is used to discuss the propagation of helicon
waves and the diffusional motion of a vortex line.
The results are discussed in light of the rather puzzling sign
change of the Hall effect which has been observed in the mixed state of the
high temperature superconductors.

\pacs 74.60.Ge, 74.40.+k


\noindent To be published in {\it Physical Review B}.

\endtopmatter

\head{\bf I. INTRODUCTION}
\taghead{1.}

The study of vortex motion in type-II superconductors continues to
attract the attention of theorists and experimentalists alike,
due in part to the rather unusual mixed state transport properties of the high
temperature superconductors.  One of the more vexing of
these properties is the anomalous behavior of the Hall effect,
which is observed to change sign in the superconducting mixed
state.$^{1-6}$\rlap{\phantom{\refto{iye89,artemenko89,hagen90,
hagen91,chien91,luo92}}}
Such a behavior is not expected within the standard models of
vortex motion in superconductors, the Bardeen-Stephen model\refto{bardeen65}
and the Nozi\`eres-Vinen model.\refto{nozieres66,vinen67} Indeed, even the low
temperature superconductors exhibited a range of behaviors not in
accord with either of the above models, including a sign change\refto{noto76};
for a review of the
early work, we refer the reader to the article by Kim and Stephen.\refto{kim69}
It is the Hall effect data which
constrains models of vortex motion in superconductors, and
which represents the greatest challenge to the theorist.
This paper is an attempt at understanding vortex motion and the Hall effect in
type-II superconductors starting from the time dependent Ginzburg-Landau
equations.

Before discussing the new results contained here, we first briefly review the
phenomenological theories of vortex motion [for a more critical assessment,
see Refs.~(\cite{vinen67}) and (\cite{kim69})].
In the Bardeen-Stephen (BS) model\refto{bardeen65} of vortex motion,
it is assumed that the vortex may be modeled as a normal core of radius the
coherence length.  If the applied transport current is
$\J_{t}=e^{*} n_{s} \vsi$, with $n_{s}$ the superfluid density (the density of
Cooper pairs), $e^{*}$ the charge of a pair (which we will take to be
positive), and $\vsi$ the uniform superfluid velocity far from the vortex,
then the Lorentz force per unit length acting upon an individual vortex is
$\F= \phi_{0}\J_{t}\times\ez$,  where $\ez$ is a unit vector which
points in the direction of the magnetic field and $\phi_{0}=h/e^{*}$ is
the flux quantum (we will take $c=1$ in this paper).
This is balanced by a viscous drag force (per unit length) $\f = -\eta \vl$,
where $\vl$ is the velocity of the vortex line.  This drag is due to the
dissipation which occurs in the normal core of the vortex,
so that $\eta\propto \sigxx$, with $\sigxx$ the longitudinal normal state
conductivity.  By balancing these two forces, $\F+\f=0$, we find
$$
\phi_{0}\J_{t}\times\ez = \eta \vl.
\eqno(bs1)
$$
Josephson\refto{josephson65} demonstrated that the motion of the
magnetic flux produces an electric field, given by Faraday's law,
$$
\langle\E\rangle = - \vl \times \B,
\eqno(faraday)
$$
where $\langle\E\rangle$ is the spatially averaged electric field and
$\B$ is the magnetic induction field.
Assuming that $\B  = B \ez$, and combining
Eqs.~\(bs1) and \(faraday), we then have
$$
\J_{t} = \frac{\eta}{\phi_{0} B} \langle\E\rangle,
\eqno(bs2)
$$
so that the flux flow conductivity is $\sigma_{xx}= \eta/(\phi_{0} B)
\approx \sigxx (H_{c2}/B)$.  In the BS model the Hall field is
entirely due to the Hall field produced in the
normal core of the vortex; the corresponding Hall angle is equal to that of a
normal metal in a field equal to the field in the core.
More importantly, the Hall angle has the same sign as in the normal state.

The Nozi\'{e}res-Vinen (NV) model\refto{nozieres66},
on the other hand, incorporates the
hydrodynamic Magnus force on the vortex,
$$
\F=\phi_{0} n_{s} e^{*} (\vsi -\vl)\times\ez.
\eqno(nv1)
$$
The Magnus force must be balanced by viscous drag forces $\f$, so that
again $\F+\f=0$.  In the absence of any viscous drag ($\f =0$), the
vortices would simply drift with the transport current ($\vl=\vsi$);
this would lead to a perfect Hall effect and no longitudinal
resistance.  The Hall fields generated in this fashion have the same
direction as the normal state Hall fields. By making rather different
assumptions regarding the nature of the contact potential at the interface
between the superfluid and the normal core,
NV conclude that the viscous force should be of the form
$\f=-a\vsi$.  The longitudinal resistivity obtained is of the same form
as the BS result, whereas NV find that the Hall angle in the mixed state
is equal to its value at the upper critical field $H_{c2}$.  However, we
still find that the Hall angle has the same sign as in the normal state.

The conclusion is that neither of these phenomenological
models is able to explain the sign
change in the Hall angle. One difficulty is that both models are
strictly speaking only valid at $T=0$;  they are also only correct at
low magnetic inductions, and should not be applied near $H_{c2}$.
However, it seems unlikely that a finite temperature generalization,
or the inclusion of intervortex interactions, would act so as to change the
sign of the Hall angle.  A more serious difficulty is that both treatments
start
from a hydrodynamic description, with no reference to the underlying
superconducting
order parameter.  It is unclear whether the inability of the BS and NV models
to predict the sign change is due to the hydrodynamic description itself
(for instance, the implicit assumption of Galilean invariance),
or with the approximations involved in the calculation.

There have also been several attempts at a fully microscopic
calculation of the Hall effect for a single vortex,
starting from the Bogoliubov-de Gennes equations for a moving
vortex.\refto{galperin76,kopnin77}
The dissipation is provided by quasiparticles which scatter from the
time dependent potential provided by the moving vortex; this is balanced
by the Magnus force on the vortex.
The Hall effect is entirely due to the Magnus force,
so this approach is also unable to explain the sign change of the
Hall coefficient.  These calculations are limited to very pure
superconductors, and do not include band structure effects which are
important in determining the sign of the Hall effect in the normal
state.

As an alternative to the hydrodynamic and microscopic approaches,
we shall study the Hall effect starting from a time dependent
version of the Ginzburg-Landau equations.  This method is intermediate
between the hydrodynamic and microscopic approaches, in that the time
dependence of the order parameter is explicitly considered, while the effects
of the quasiparticles are lumped into an effective conductivity for the
``normal fluid.''  The scheme is to then use this model to systematically
study the motion of a single vortex.
This program has already been carried out for the
longitudinal resistivity by Schmid,\refto{schmid66} Gor'kov and
Kopnin,\refto{gorkov71,gorkov76} and Hu and Thompson.\refto{hu72}
The time dependent Ginzburg-Landau (TDGL) equations must
be generalized somewhat
in order to study the Hall effect; with this generalization, we will
show that a single vortex has the equation of motion
$$
\vsi\times\ez = \alpha_{1} \vl + \alpha_{2}\vl\times\ez,
\eqno(nv2)
$$
where $\alpha_{1}$ and $\alpha_{2}$ are functions of the
parameters which appear in the TDGL equations.  Such an equation
of motion for superconducting vortices was originally proposed by
Vinen and Warren\refto{vinen67};  a similar phenomenological model
has recently been used by Hagen {\it et al.}\refto{hagen91} to discuss
the sign change of the Hall angle in YBCO.  We see from Eq.~\(nv2) that
$\alpha_{1}$ will determine the longitudinal conductivity, while
$\alpha_{2}$ determines the Hall conductivity.  In particular,
if $\alpha_{2}<0$, then the Hall effect in the vortex state will have a
sign which is opposite to the sign of the normal state Hall effect.
Having reduced the problem to
this effective equation of motion, one can then pose the question,
``What choice of parameters leads to a Hall effect which changes sign?''

The outline of the remainder of the paper is as follows.  In Section II
the time dependent Ginzburg-Landau equations are presented and
discussed.  In Section III we derive an equation of motion for a single
vortex, starting from the time dependent Ginzburg-Landau equations.  From
this equation of motion the longitudinal and Hall conductivities are
calculated and compared to the predictions of the Bardeen-Stephen
and Nozi\'{e}res-Vinen
models.  We also recover known results for the motion of a rectilinear
vortex in a neutral superfluid in the appropriate limit.  In Section IV
we study the thermal transport properties of a single vortex, and calculate
the Nernst coefficient and the thermopower. The effects of vortex bending and
fluctuations are considered in Section V.  We will derive a  Langevin equation
for vortex motion, which we will use to study helicon waves and the
diffusive motion of the vortex center of mass.  The relevance of these results
to the anomalous transport properties of the high temperature superconductors
is discussed in Section VI.  The London acceleration equation for a
charged superfluid is derived from the TDGL equations in Appendix~A.
Numerical coefficients which enter the vortex
equation of motion are calculated using a trial solution for the order
parameter in Appendix~B.   Appendix~C is a summary of the definitions of the
transport coefficients.

\head{\bf II. TIME DEPENDENT GINZBURG-LANDAU EQUATIONS}
\taghead{2.}

\subhead{\bf A. The model}

To begin our discussion of vortex motion in superconductors, we
first need an appropriate generalization of the familiar equilibrium
Ginzburg-Landau equations to include dynamics.
Such generalizations have been the subject of intensive
study over the years, starting with Schmid's derivation of
TDGL equations.\refto{schmid66,cyrot73}
Gor'kov and \'{E}liashberg later
showed that Schmid's results were only valid in the dirty limit, and derived
a modified version of the TDGL equations which are
valid when the pair breaking is due to paramagnetic impurities.\refto{gorkov68}
These equations were further developed to include the effects of
electron-phonon scattering on the order parameter
relaxation.$^{21-23}$\rlap{\phantom{\refto{kramer78,schon79,watts81}}}
While there are some important
consequences of these generalizations, they result in more complicated and
cumbersome dynamic equations; we will therefore adopt the TDGL equations
originally
proposed by Schmid (with minor modifications) as the prototypical
equations of motion.  Most of the
results in this paper may be generalized in a straightforward, if not
tedious, manner to the other more complicated dynamical equations.

Our equation of motion for the superconducting order parameter $\psi(\r,t)$
is
$$
\hbar (\partial_{t} + i\frac{\tilde{\mu}}{\hbar})\psi = -\Gamma
          \frac{\delta {\cal H}}{\delta \psi^{*}},
\eqno(ham1)
$$
with the Hamiltonian
$$
{\cal H} = \int d^{3}r \left[ \frac{\hbar^{2}}{2m} \left|
     ( \nabla - i\frac{e^{*}}{\hbar} \A )\psi\right|^{2} + a(T) |\psi|^{2}
      + \frac{b}{2}|\psi|^{4}  + \frac{1}{8\pi} (\nabla\times \A)^{2} \right].
\eqno(ham2)
$$
In the above equations, $\A$ is the vector potential with
$\h=\nabla\times\A$ the microscopic magnetic induction field,
$\B = \langle \h\rangle$ s the induction field, $m$ is the effective mass of
a Cooper pair, $e^{*}=2e$ is the charge of a Cooper pair (we take $e^{*}$ to be
positive), $a(T)=a_{0}(T/T_{c} -1)$, and $\Gamma=\Gamma_{1}+ i\Gamma_{2}$ is a
complex dimensionless relaxation rate.  For anisotropic superconductors
(such as the high temperature superconductors)
we would need to allow for an effective mass tensor $m_{ij}$;  in order to
simplify the discussion we shall assume that $m_{ij}=m\delta_{ij}$.
This assumption will be relaxed when vortex bending is considered in Sec. V.
As long as $\Gamma_{1} >0$, this
equation of motion relaxes to the correct equilibrium Ginzburg-Landau
equation.
The total chemical potential $\tilde{\mu}$ is given by
$$
\tilde{\mu} = \mu + e^{*}\Phi + \frac{\delta {\cal H}}{\delta n_{s}},
\eqno(ham3)
$$
where $\mu$ is the chemical potential, $\Phi$ is the
electric potential, $n_{s} = |\psi|^{2}$ is the superfluid density,
and $\delta {\cal H}/\delta n_{s}$ is the kinetic energy of the superfluid.
The last contribution is often neglected, although it is essential if one
desires a Galilean invariant equation of motion.\refto{abrahams66} If we set
$(\delta {\cal H}/\delta n_{s})\psi \approx \delta {\cal H}/\delta\psi^{*}$,
then we can rewrite Eq.~\(ham1) as
$$
\hbar (\partial_{t} + i \frac{1}{\hbar}\mu + i\frac{e^{*}}{\hbar}\Phi)\psi
  = -(\Gamma + i) \frac{\delta {\cal H}}{\delta \psi^{*}}.
\eqno(ham4)
$$
Similar equations
have been used to study the hydrodynamics of superfluid ${\rm He}^{4}$
near the $\lambda$-transition.$^{25-27}$\rlap{\phantom{\refto{khalatnikov,
onuki83a,onuki83b}}} By defining a dimensionless order parameter relaxation
time
$$
\gamma\equiv\gamma_{1} + i\gamma_{2}= \frac{\Gamma_{1} - i(1+\Gamma_{2})}
                      {\Gamma_{1}^{2} + (1+\Gamma_{2})^{2}},
\eqno(gamma)
$$
our order parameter equation of motion becomes\refto{schmid66}
$$
\hbar \gamma(\partial_{t} + i\frac{e^{*}}{\hbar}\tilde{\Phi})\psi
= \frac{\hbar^{2}}{2m}(\nabla - i\frac{e^{*}}{\hbar}\A)^{2}\psi
+|a|\psi - b|\psi|^{2} \psi,
\eqno(gl1)
$$
where $\tilde{\Phi}= \Phi + \mu/e^{*}$.  The difference between $\tilde{\Phi}$
and $\Phi$ is generally small,\refto{schmid66} and we shall neglect this
difference in what follows.

By choosing the complex relaxation time $\gamma$ appropriately, we can
consider a variety of different models.
If $\Gamma_{2}=0$,  order parameter
equation of motion leads to the London acceleration equation for the
superfluid velocity, as shown in Appendix A.  If $\Gamma_{2} = -1$ (so that
$\gamma_{1}=\Gamma_{1}^{-1}$ and $\gamma_{2}=0$), then we have the TDGL
originally derived by Schmid.\refto{schmid66} If $\Gamma_{1}=\Gamma_{2}=0$
(so that $\gamma_{1}=0$ and $\gamma_{2}=-1$), then we have the
Gross-Pitaevskii equation\refto{gross61} (often called the nonlinear
Schr\"{o}dinger equation) for a charged superfluid at zero
temperature.  More generally, such an imaginary part of the relaxation
time is generated in renormalized theories of the critical
dynamics of neutral superfluids.\refto{dedominicis78}
A microscopic derivation of the TDGL by Fukuyama, Ebisawa, and
Tsuzuki,\refto{fukuyama71} leads to a value of $\gi$ that depends
on details of the
band structure of the material, and  is generally
proportional to the derivative of the density of states at the fermi
energy, $N'(\epsilon_{F})$.  In what follows we will consider $\gamma$
to be arbitrary, and after having derived an equation of motion for a vortex,
we can then consider specific models for $\gamma$.

We also require an equation of motion for the the vector potential, which
is just Amp\`{e}re's Law:
$$
\nabla\times\nabla\times \A = 4\pi(\Jn + \Js),
\eqno(gl2)
$$
so that $\nabla\cdot(\Jn + \Js)=0$.  The supercurrent $\J_{s}$ is given by
$$
\Js=\frac{\hbar e^{*}}{2 m i} (\psi^{*}\nabla\psi - \psi\nabla\psi^*)
 - \frac{(e^{*})^{2}}{m}|\psi|^{2} \A,
\eqno(super)
$$
while the normal current $\J_{n}$ is given by
$$\eqalign{
\Jn = & \sig\cdot \E\cr
     &= \sig\cdot (-\nabla\Phi - \partial_{t}\A)\cr}
\eqno(normal)
$$
with $\sig$ the normal state conductivity tensor,
$$
\sig = \left(\matrix{\sigxx & \sigxy \cr
                     \sigyx & \sigxx \cr}\right).
\eqno(sigma)
$$
The Onsager relations and rotational symmetry imply that
$\sigyx (\H)=-\sigxy (\H)$, so that the conductivity
tensor may be decomposed into a diagonal piece and an antisymmetric piece.
The longitudinal normal state conductivity $\sigxx$ is generally a
weak function of the magnetic field, and we will consider it to be field
independent; the  Hall conductivity in the low field limit is of the
form $\sigxy (H)= \omega_{c} \tau \sigxx$, with
$\omega_{c}= e H/m$ the cyclotron frequency and $\tau$ the scattering
time.  Since in what follows we will be considering inhomogeneous magnetic
fields, the normal state Hall conductivity is generally a function of
position.
In equilibrium the electric field is zero, and the TDGL equations reduce to the
familiar equilibrium Ginzburg-Landau equations.\refto{fetter}

The  new features in Eqs.~\(gl1) and \(gl2) are the imaginary part
in the order parameter relaxation time, $\gi$,  and  the Hall conductivity
for the normal fluid, $\sigxy$.  These terms are crucial for understanding the
Hall effect in the mixed state.\refto{ullah91,dorsey92}
To see why, notice that if $\gi$ and $\sigxy$ are both zero the
TDGL equations have an important {\it particle-hole symmetry}: the
equations are invariant under the simultaneous transformations
$\psi\rightarrow\psi^{*}$, $\Phi\rightarrow -\Phi$, and
 $\A\rightarrow -\A$.  Under this transformation the total current
$\J=\J_{n}+\J_{s}$ changes sign, as does the electric field.
If we define the total conductivity tensor $\sigma_{\mu\nu}(\H)$
in terms of the spatially averaged current and electric field as
$J_{\mu}(\H)= \sigma_{\mu\nu}(\H) E_{\nu}(\H)$, then
upon reversing the magnetic field,
$J_{\mu}(-\H)= \sigma_{\mu\nu}(-\H) E_{\nu}(-\H)$; but because of the
particle-hole
symmetry, $J_{\mu}(-\H)=-J_{\mu}(\H)$ and $ E_{\nu}(-\H)=- E_{\nu}(\H)$,
so that $ \sigma_{\mu\nu}(-\H)= \sigma_{\mu\nu}(\H)$; under these conditions
the conductivity tensor is even in the magnetic field.  However,
we know from the Onsager reciprocity relations\refto{callen} that
in general $\sigma_{\mu\nu}(\H)=\sigma_{\nu\mu}(-\H)$. Rotational
invariance in the plane perpendicular to the applied field requires that
the off diagonal components of the conductivity tensor satisfy
$\sigma_{\mu\nu}(\H)=-\sigma_{\nu\mu}(\H)$; when combined with the Onsager
relations, we find that the off-diagonal components of the conductivity
tensor must be odd in the magnetic field for  a rotationally invariant system.
We therefore conclude that if
$\gi=\sigxy=0$, then the Hall conductivity
$\sigma_{xy}(\H)\equiv 0$.  If {\it either}
of these  quantities is nonzero, then the particle-hole symmetry
is destroyed, and there will be a nonvanishing Hall conductivity.
The term $\gi$ produces a Hall effect due to the Magnus force on the vortex,
while $\sigxy$ produces a Hall effect due the transverse response of the
normal fluid to the electric fields generated  in the vortex core.

\subhead{\bf B. Dimensionless units}

In order to facilitate the calculations it is helpful to recast
Eqs.~\(gl1)--\(sigma) into dimensionless units:
$$
\eqalign{
& \r = \lambda \r';\qquad t=(\hbar/|a|)t';\qquad \psi=(|a|/b)^{1/2} \psi';\cr
&\A = \sqrt{2} H_{c}\lambda \A';\qquad \Phi = (e^{*}/|a|) \Phi';
\qquad \sigma = (2m/\hbar) (1/4\pi\kappa^{2})\sigma',\cr}
\eqno(dims)
$$
where the magnetic penetration depth $\lambda=[mb/4\pi (e^{*})^{2} |a|]^{1/2}$,
the coherence length $\xi = \hbar/(2m|a|)^{1/2}$, the Ginzburg-Landau
parameter $\kappa=\lambda/\xi$,
and the thermodynamic critical field $H_{c}^{2}=4\pi |a|^{2}/b$.
In these units the
equations become (we will henceforth drop the primes on the dimensionless
quantities)
$$
(\gr + i\gi)(\partial_{t} + i\Phi)\psi
= (\frac{\nabla}{\kappa} - i\A)^{2}\psi
+\psi - |\psi|^{2} \psi,
\eqno(nodim)
$$
$$
\nabla\times\nabla\times \A = \Jn + \Js,
\eqno(nodim1)
$$
$$
\Jn = \sig\cdot (-\frac{1}{\kappa}\nabla\Phi - \partial_{t}\A),
\eqno(nodim2)
$$
$$
\Js=\frac{1}{2 \kappa i} (\psi^{*}\nabla\psi - \psi\nabla\psi^*)
 - |\psi|^{2} \A.
\eqno(nodim3)
$$
The ``$\cdot$'' in Eq.~\(nodim2) indicates a tensor product.
The superfluid velocity, in conventional units, is $\vs=\Js/e^{*}|\psi|^{2}$.
In our dimensionless units this becomes
$$
\Js = \frac{\kappa}{2} f^{2} \vs.
\eqno(nodim4)
$$

\subhead{\bf C.  Simplification of the TDGL equations}

First, we rewrite the complex order parameter in terms of an amplitude
and a phase, $\psi(\r,t)=f(\r,t) \exp [i\chi(\r,t)]$. (Note that a
moving vortex does not possess cylindrical symmetry, so that the phase
variable $\chi$ is equal to the angular variable $\theta$ only near the
center of the vortex.)
In terms of the gauge invariant quantities
$\Q\equiv \A - \nabla\chi/\kappa$ and $ P  \equiv \Phi + \partial_{t} \chi$,
the magnetic and electric fields are
$$
\h=\nabla\times\Q,
\eqno(magnetic)
$$
$$
\E=-\frac{1}{\kappa}\nabla P - \partial_{t} \Q.
\eqno(electric)
$$
The real part of Eq.~\(nodim) is
$$
\gr\partial_{t}f  -\gi P f
= \frac{1}{\kappa^{2}} \nabla^{2} f - Q^{2} f + f - f^{3},
\eqno(real)
$$
while the imaginary part is
$$
\gi \partial_t f + \gr P f + \frac{1}{\kappa} f \nabla\cdot \Q
        + \frac{2}{\kappa} \Q \cdot \nabla f = 0,
\eqno(imag1)
$$
and Eqs.~\(nodim1)-\(nodim3) become
$$
\nabla\times\nabla\times \Q = \sig\cdot(-\frac{1}{\kappa}\nabla P
 - \partial_{t} \Q) - f^{2}\Q .
\eqno(vectpot)
$$
To derive an equation for the potential $P$,
first multiply Eq.~\(imag1) by $f$:
$$
\gr P f^{2} + \frac{1}{\kappa}\nabla\cdot(f^{2}\Q)
+ \gi f\partial_{t}f = 0.
\eqno(imag2)
$$
Next, use the fact that $\nabla\cdot (\J_{s} + \J_{n}) =0$ to obtain
$$
\nabla\cdot(\sig \cdot\E) - \nabla\cdot(f^{2}\Q) = 0.
\eqno(div)
$$
Finally, combining Eqs.~\(imag2) and \(div) we obtain
$$
\frac{1}{\kappa}\nabla\cdot[\sig\cdot(-\frac{1}{\kappa}\nabla  P
 - \partial_{t}\Q)] + \gr f^{2} P + \gi f\partial_{t}f = 0.
\eqno(chempot1)
$$
The remainder of the paper is devoted to solving Eqs.~\(real), \(vectpot),
and \(chempot1) for a moving vortex.  If we set $\gi$ and
$\sigma_{xy}^{(n)}=0$, then our
equations are identical to those studied by Schmid,\refto{schmid66}
Hu and Thompson,\refto{hu72} and Gor'kov and Kopnin,\refto{gorkov71}
in the context of the viscous motion of a single vortex. These equations
are therefore a generalization which allows for the
possiblility of a nonzero Hall conductivity.

\head{\bf III. VORTEX EQUATION OF MOTION IN THE LIMIT $B\ll H_{c2}$}
\taghead{3.}

Since the full nonlinear TDGL equations are complicated, we want
to focus attention on the motion of the vortices, as they are the
``elementary excitations'' of the mixed state.
In this paper we will derive an equation of motion for a single
vortex; we therefore consider magnetic fields which are slightly
above the lower critical field $H_{c1}$.
The vortex motion, and the concomitant motion of magnetic flux, lead
to dissipation in the mixed state of type-II superconductors.
Several approaches have been adopted in order to study the
vortex motion.  Schmid\refto{schmid66} constructed a dissipation
functional starting from the TDGL, and from energy balance arguments
he was able to calculate the flux flow conductivity in terms of the
parameters of the TDGL equations.   Hu and Thompson\refto{hu72,hu73}
also used energy balance arguments to calculate the flux flow conductivity,
but they included important backflow
contributions which had been neglected by Schmid.
This method is intuitive and easily implemented, but is insufficient for
our purposes as Hall fields are nondissipative.  Instead, we will
use the method developed by Gor'kov and Kopnin\refto{gorkov71,gorkov76}
in their study of flux flow; this method has also been used to study
the mutual friction of vortices in superfluid HeII near the
$\lambda$ point.\refto{onuki83a,onuki83b,sonin81,sonin87} There are essentially
three steps to the calculation.  We first assume that the vortices
translate uniformly, so that the order parameter, vector potential, and
chemical potential are functions of the quantity $\r- \vl t$, where
$\vl$ is the vortex line velocity.  Next, we assume that these
quantities may be expanded in powers of $\vl$.  The terms of
$O(1)$ are simply the equilibrium Ginzburg-Landau equations, while the
$O({\rm v_{L}})$ equations are a set of linear, inhomogeneous differential
equations. These equations will only have solutions for particular values
of $\vl$.  Therefore, the final step is to derive a ``solvability condition''
for $\vl$, which is tantamount to deriving an equation of motion for the
vortices.  This equation of motion, along with Faraday's law for the
moving vortices,\refto{josephson65}
$\langle\E\rangle =- \vl \times \B $,
lead to the longitudinal and Hall conductivities.

\subhead{\bf A. Derivation of the solvability condition}

First, we assume that $f$, $\Q$, and $P$
are only functions of ${\bf r} - \vl t$.  Therefore we replace all time
derivatives in Eqs.~\(real), \(vectpot), and \(chempot1)
by $-\vl\cdot\nabla$, and obtain the following set of equations:
$$
-\gr \vl\cdot\nabla f -\gi P f = \frac{1}{\kappa^{2}}\nabla^{2}f
- Q^{2}f + f - f^{3},
\eqno(linear2)
$$
$$
\nabla\times\nabla\times \Q = \sig\cdot[-\frac{1}{\kappa}\nabla P
 + (\vl\cdot\nabla) \Q] - f^{2}\Q,
\eqno(vectpot2)
$$
$$
\frac{1}{\kappa}\nabla\cdot\{\sig\cdot[-\frac{1}{\kappa}\nabla P
 + (\vl\cdot\nabla)\Q]\} + \gr f^{2} P - \gi f \vl\cdot\nabla f = 0,
\eqno(linear1)
$$
where $P = \Phi -\vl\cdot\nabla\chi$.

Next, we expand all quantities in powers of the vortex velocity;
$f= f_{0} + f_{1}$, $\Q = \Q_{0} + \Q_{1}$,
where $f_{1}$ and $\Q_{1}$ are $O({\rm v}_{L})$.
Note that $P$ is $O({\rm v}_{L})$, since the electric
field vanishes in equilibrium.  The $O(1)$ equations are simply the
equilibrium Ginzburg-Landau equations,
$$
\frac{1}{\kappa^{2}}\nabla^{2}f_{0} - Q_{0}^{2}f_{0} + f_{0} - f_{0}^{3}=0,
\eqno(order1)
$$
$$
\nabla\times\nabla\times \Q_{0} + f_{0}^{2} \Q_{0} = 0,
\eqno(order2)
$$
with the equilibrium supercurrent given by
$$
\J_{0} = -f_{0}^{2} \Q_{0}.
\eqno(current1)
$$
Next, we need the $O({\rm v}_{L})$ equations.  In terms of the quantities
$f_{v}\equiv \vl\cdot\nabla f_{0}$, $\Q_{v}\equiv(\vl\cdot\nabla) \Q_{0}$,
these are
$$
\frac{1}{\kappa^{2}}\nabla^{2}f_{1} - Q_{0}^{2}f_{1} - 2f_{0}\Q_{0}\cdot\Q_{1}
+ f_{1} - 3f_{0}^{2}f_{1} + \gi  Pf_{0} = -\gr f_{v},
\eqno(order21)
$$
$$
\nabla\times\nabla\times \Q_{1} + f_{0}^{2}\Q_{1} + 2 f_{0}f_{1} \Q_{0}
 + \frac{1}{\kappa} \sig\cdot \nabla P = \sig\cdot \Q_{v},
\eqno(order23)
$$
$$
-\frac{1}{\kappa^{2}}\nabla\cdot(\sig\cdot\nabla P) + \gr f_{0}^{2} P
= \gi f_{0} f_{v} - \frac{1}{\kappa} \nabla\cdot(\sig\cdot \Q_{v}),
\eqno(order22)
$$
with the current $\J_{1}= \J_{1s} + \J_{1n}$, where
$$
\J_{1s} = - f_{0}^{2}\Q_{1} - 2f_{0}f_{1} \Q_{0},
\eqno(order24)
$$
$$
\J_{1n}= \sig\cdot \left(-\frac{1}{\kappa} \nabla P + \sig\cdot \Q_{v}\right).
\eqno(order25)
$$
Far from the center of the vortex $\J_{1n}\rightarrow 0$, and
$\J_{1s}$ is equal to the applied transport current $\J_{t}$.
Eqs.~\(order21)-\(order22) are a set
of inhomogeneous linear equations which must be solved in order to
determine the vortex velocity.

We next derive a solvability condition for the linear inhomogeneous
equations which will determine the vortex velocity.  Following
Gor'kov and Kopnin\refto{gorkov76} we note that the time {\it independent}
Ginzburg-Landau equations possess a translational invariance, so that
if $f_{0}({\bf r})$ and $\Q_{0}(\r)$  are solutions, so are
$f_{0}({\bf r}+{\bf d})$ and $\Q_{0}(\r+\d)$,
with ${\bf d}$ an arbitrary translation vector.\refto{rajaraman}
If ${\bf d}$ is an infinitesimal translation, then we have
$f_{0}(\r+\d)= f_{0}(\r) +\d\cdot\nabla f_{0}(\r)+\ldots\ $,
so that the quantities
$f_{d}\equiv \d\cdot\nabla f_{0}$ and $\Q_{d}\equiv (\d\cdot\nabla) \Q_{0}$
will solve the linear equations  \(order21) and \(order23)
without the inhomogeneous terms on the right hand side
and with $P=0$:
$$
\frac{1}{\kappa^{2}}\nabla^{2}f_{d} - Q_{0}^{2}f_{d} - 2f_{0}\Q_{0}\cdot\Q_{d}
+ f_{d} - 3f_{0}^{2}f_{d}  = 0,
\eqno(d1)
$$
$$
\nabla\times\nabla\times \Q_{d} + f_{0}^{2}\Q_{d} + 2 f_{0}f_{d} \Q_{0}=0,
\eqno(d2)
$$
with the current
$$\eqalign{
\J_{d}& \equiv (\d\cdot\nabla) \J_{0}\cr
      &=-2f_{0}f_{d} \Q_{0} - f_{0}^{2}\Q_{d}.\cr}
\eqno(jd)
$$
To derive the solvability condition, we
multiply Eq.~\(order21) by $f_{d}$ and integrate over a cylindrical volume.
We then integrate the term $f_{d}\nabla^{2}f_{1}$ by parts twice,
discarding the surface terms, and we use Eq.~\(d1) for
$f_{d}$ to eliminate the nonlinear terms in the integral.  We finally
obtain
$$
\int d^{2}r \left( 2f_{0}f_{1} \Q_{0}\cdot\Q_{d}
- 2f_{0}f_{d} \Q_{0}\cdot\Q_{1} + \gr f_{v}f_{d} + \gi P f_{0}f_{d}
\right) = 0.
\eqno(solve)
$$
Using Eq.~\(order24) for $\J_{1s}$ and Eq.~\(jd) for $\J_{d}$,
we may express Eq.~\(solve) in terms of the currents as
$$
\int d^{2}r \left( \J_{d}\cdot \Q_{1} - \J_{1s}\cdot\Q_{d}\right)
  = -\int d^{2}r \left( \gr f_{v}f_{d} + \gi P f_{0}f_{d}\right).
\eqno(solve2)
$$
Further simplification occurs in the large $\kappa$ limit,\refto{gorkov76}
where $\Q_{d}\approx - \nabla \chi_{d}/\kappa$,
$\Q_{1}\approx - \nabla \chi_{1}/\kappa$. Inserting these expressions into
the left hand side of Eq.~\(solve2), we have
$$\eqalign{
\int d^{2}r \left( \J_{d}\cdot \Q_{1} - \J_{1s}\cdot\Q_{d}\right)
    &= \frac{1}{\kappa}\int d^{2}r \left( \J_{1s}\cdot \nabla\chi_{d}
           - \J_{d}\cdot\nabla\chi_{1} \right) \cr
    &=\frac{1}{\kappa}\int d^{2}r\left[ \nabla\cdot (\J_{1s}\chi_{d})
        - \nabla\cdot (\J_{d}\chi_{1})\right]
          -\frac{1}{\kappa}\int d^{2}r \chi_{d}\nabla\cdot \J_{1s},\cr}
\eqno(surface)
$$
where we have used the fact that $\nabla\cdot\J_{d} = 0$.  The first
integral on the right hand side of Eq.~\(surface) may be converted into
a surface integral. The second term may be rewritten using
$\nabla\cdot \J_{1s}= -\nabla\cdot \J_{1n}$, and then by
using Eq.~\(order22) we find
$$\eqalign{
\frac{1}{\kappa}\int d^{2}r \chi_{d}\nabla\cdot \J_{1s}
        & = -\frac{1}{\kappa}\int d^{2}r \chi_{d}\nabla\cdot \J_{1n} \cr
        &=\int d^{2}r \chi_{d} (\gr f_{0}^{2} P -\gi f_{0} f_{v}). \cr}
\eqno(surface2)
$$
Combining Eqs.~\(solve2), \(surface), and \(surface2), we have
$$
\frac{1}{\kappa}\int d{\bf S}\cdot(\J_{1s}\chi_{d} - \J_{d}\chi_{1})=
     -\int d^{2}r \left( \gr f_{v}f_{d} - \gr \chi_{d} f_{0}^{2} P
        + \gi P f_{0}f_{d} + \gi\chi_{d} f_{0} f_{v} \right).
\eqno(solve3)
$$
This is our solvability condition for steady vortex
motion, which is exact to linear order in the vortex velocity.
If this solvability condition does not hold, then the linear inhomogeneous
equations have no solutions, and steady vortex motion is impossible.
The remainder of this Section is devoted to evaluating Eq.~\(solve3).

\subhead{\bf B. Coordinate system and core fields}

The coordinate system which will be used for evaluating Eq.~\(solve3)
is defined in Fig. 1.  The applied transport current $\J_{t}$ is assumed to
be in the $x$-direction; the magnetic field is in the $z$-direction (out of
the page); the vortex moves at an angle $\theta_{H}$ with respect
to the $-y$-direction, so that the averaged electric field $\langle \E\rangle$
makes an angle $\theta_{H}$ with the $x$-axis.  We will use $(r,\theta,z)$
as our cylindrical coordinates, with unit vectors $\er$, $\etheta$, and $\ez$
respectively.  The displacement vector $\d$ makes an angle $\phi$
with respect to the $x$-axis. The explicit forms of $\J_{t}$, $\vl$, and
$\d$ in this cylindrical coordinate system are
$$
\J_{t} = J_{t} [\cos\theta\ \er - \sin\theta\ \etheta],
\eqno(jt)
$$
$$
\vl = -v_{L} [\sin(\theta-\theta_{H})\ \er + \cos(\theta-\theta_{H})\ \etheta],
\eqno(vl)
$$
$$
\d = d[\cos(\theta -\phi)\ \er - \sin(\theta-\phi)\ \etheta].
\eqno(d)
$$
In this coordinate system, we have $\vl\cdot\d=-v_{L} d \sin(\phi-\theta_{H})$
and $(\vl\times\ez)\cdot\d = - v_{L} d \cos (\phi-\theta_{H})$.

Before evaluating the solvability condition, we first want to simplify the
equations for the order parameter, vector potential, and scalar potential,
in order to bring them into a more manageable form. The
equilibrium order parameter $f_{0}(r)$ and vector potential
$\Q_{0}(r)= Q_{0}(r)\ \etheta$  satisfy Eqs.~\(order1) and \(order2),
which in the cylindrical coordinates defined above become
$$
\frac{1}{\kappa^2}\frac{1}{r}\frac{d}{dr}\left( r \frac{d f_{0}}{dr}\right)
  -Q_{0}^{2} f_{0}     + f_{0} - f_{0}^{3} = 0 ,
\eqno(neworder1)
$$
$$
\frac{d}{dr}\frac{1}{r}\frac{d(r Q_{0})}{dr}
 -  f_{0}^{2} Q_{0} = 0,
\eqno(newvectpot)
$$
with $Q_{0}(r)\sim -1/(\kappa r)$ as $r\rightarrow 0$.

The equation for the gauge invariant scalar potential is rather more
complicated.  It can be simplified somewhat if we ignore spatial derivatives of
the normal state Hall conductivity, which should be small (especially near
the center of the vortex). Then $\nabla\cdot(\sig\cdot\nabla P)
 = \sigxx\nabla^{2} P$. Using the coordinate system defined above and
$\h_{0}=\nabla\times\Q_{0}$, Eq.~\(order22) then becomes
$$
\frac{\sigxx}{\kappa^{2}}\nabla^{2} P - \gr f_{0}^{2}P
   = \left[ \gi f_{0}\frac{\partial f_{0}}{\partial r} - \frac{\sigxy}{\kappa}
     \frac{\partial h_{0}}{\partial r} \right] v_{L} \sin (\theta-\theta_{H}).
\eqno(scalar1)
$$
Now we notice that as $r\rightarrow 0$,
$P\approx -\vl\cdot\nabla\theta=v_{L}\cos(\theta-\theta_{H})/r$.  We therefore
decompose $P(\r)$ as
$$
P(\r) = v_{L} [p_{1}(r)\cos(\theta-\theta_{H})
       + p_{2}(r)\sin(\theta-\theta_{H})].
\eqno(scalar2)
$$
The contribution $p_{1}(r)$ satisfies a homogeneous equation
$$
\frac{\sigxx}{\kappa^2}\frac{d}{dr}\frac{1}{r}\frac{d(rp_{1})}{dr}
 -  \gr f_{0}^{2} p_{1} = 0,
\eqno(chem1)
$$
with the boundary condition $p_{1}(r)\sim 1/r$ as $r\rightarrow 0$.
The contribution $p_{2}$ is due to the particle-hole asymmetry, and
satisfies an inhomogeneous equation
$$
\frac{\sigxx}{\kappa^2}\frac{d}{dr}\frac{1}{r}\frac{d(rp_{2})}{dr}
 -  \gr f_{0}^{2} p_{2} = \gi f_{0}\frac{d f_{0}}{d r}
    - \frac{\sigxy}{\kappa} \frac{d h_{0}}{d r},
\eqno(chem2)
$$
with the boundary condition $p_{2}(r=0)= 0$.  We also have
$p_{1}(r)$, $p_{2}(r)\rightarrow 0$ as $r\rightarrow \infty$.
This means that the homogeneous contribution to $p_{2}(r)$
is identically zero, so that $p_{2}(r)$ is $O(\gi,\sigxy)$.
Note that the equations for $p_{1}$ and $p_{2}$ are decoupled; if we had
included the terms involving spatial gradients of the Hall conductivity
then these two equations would be coupled together, an inessential
complication for our purposes.

It is now possible to determine the electric field at the core of the
vortex, $\E(0)$. First, notice that as $r\rightarrow 0$, the scalar and vector
potentials have the following behaviors:
$$
Q_{0}(r) \approx -\frac{1}{\kappa r} + \half h_{0}(0) r,
\eqno(q0)
$$
$$
p_{1}(r)\approx \frac{1}{r} - p_{1}^{(1)} r,
           \qquad p_{2}(r)\approx - p_{2}^{(1)} r,
\eqno(p0)
$$
where $h_{0}(0)$ is the magnetic field at the center of the vortex, and
where $p_{1}^{(1)}$ and $p_{2}^{(1)}$ are constants which are determined
from the solution of Eqs.~\(chem1) and \(chem2). In terms of these constants,
the electric field at the center of the vortex is
$$
\eqalign{
\E(0) & = \left[ -\frac{1}{\kappa} \nabla P +
          (\vl\cdot\nabla)\Q_{0}\right]_{r=0} \cr
      & = -\frac{p_{2}^{(1)}}{\kappa} \vl - \left[ \frac{p_{1}^{(1)}}{\kappa} +
          \half h_{0}(0)\right] \vl\times\ez,\cr}
\eqno(e0)
$$
where we have used Eqs.~\(scalar2), \(q0), and \(p0).
Explicit expressions for $p_{1}^{(1)}$ and $p_{2}^{(1)}$  may be found in
Appendix B.   The electric field at
the core of the vortex is not parallel to the averaged electric field,
$\langle \E\rangle$, except for the particle-hole symmetric case
($p_{2}^{(1)}=0$).

\subhead{\bf C. Evaluation of the solvability condition}

We start by evaluating  the left
hand side of Eq.~\(solve3).
The surface integral can be expressed in terms of the applied
transport current at the
boundaries, since at the boundaries $\J_{1s}(r=\infty,\theta) = \J_{t}$,
$\J_{d}\cdot\er=d\sin(\theta-\phi)/(\kappa r^{2})$,
$\chi_{d}\equiv \d\cdot\nabla \theta =-d\sin(\theta-\phi)/r$ and
$\chi_{1} = \kappa J_{t} r\cos\theta$.  Substituting these expressions into
the left hand side of Eq.~\(solve3), and performing the remaining
angular integral, we find
$$
\frac{1}{\kappa} \int d{\bf S}\cdot\left[ \J_{1s} \chi_{d}
   - \J_{d}\chi_{1} \right] = -\frac{2\pi}{\kappa} (\J_{t}\times \ez) \cdot \d.
\eqno(solve7)
$$
This term represents the driving force on the vortex, due to the applied
transport current.  It is balanced by the viscous forces on the right
hand side of the solvability condition, Eq.~\(solve3).

The next step is to evaluate the right hand side of Eq.~\(solve3).
The first term is
$$
\gr\int d^{2}rf_{v}f_{d}=  \pi \vl\cdot\d \gr \int_{0}^{\infty}
              ( f_{0}')^{2} r \,dr,
\eqno(right1)
$$
where the prime denotes a derivative with respect to $r$.
The second term is
$$
\gr\int d^{2}r \chi_{d} f_{0}^{2}P = -\pi \vl\cdot\d
\gr\int_{0}^{\infty} f_{0}^{2} p_{1} \,dr + \pi (\vl\times\ez)\cdot\d
\gr\int_{0}^{\infty}  f_{0}^{2}p_{2} \,dr.
\eqno(right2)
$$
The third term is
$$
\gi\int d^{2}r P f_{0} f_{d} =- \pi (\vl\times \ez)\cdot \d
 \frac{\gi}{2} \int_{0}^{\infty}  (f_{0}^{2})' p_{1} r\, dr
  - \pi \vl\cdot\d
    \frac{\gi}{2} \int_{0}^{\infty} (f_{0}^{2})'  p_{2} r \,dr.
\eqno(right3)
$$
For the fourth term we have
$$
\gi\int d^{2}r \chi_{d} f_{0} f_{v} =-\frac{\pi}{2} \gi (\vl\times\ez)\cdot\d.
\eqno(right4)
$$
Collecting together the various terms on the right hand side,
Eqs.~\(right1)-\(right4), equating them with the driving force on the
left hand side, Eq.~\(solve7), and recalling that the
displacement vector $\d$ is arbitrary, we obtain the following
equation of motion for the vortex:
$$
 \J_{t}\times \ez = \frac{\alpha_{1} \kappa}{2} \vl
     + \frac{\alpha_{2} \kappa}{2} \vl \times \ez,
\eqno(motion)
$$
where the constants $\alpha_{1}$ and $\alpha_{2}$ are given by
$$
\alpha_{1}  = \gr \int_{0}^{\infty} (f_{0}')^{2} r \,dr
              + \gr \int_{0}^{\infty}f_{0}^{2} p_{1} \,dr
              - \frac{\gi}{2}\int_{0}^{\infty} (f_{0}^{2})'  p_{2} r \,dr,
\eqno(alpha1)
$$
$$
\alpha_{2}  =- \frac{\gi}{2}\int_{0}^{\infty} (f_{0}^{2})'p_{1} r \,dr
       -\frac{\gi}{2} -\gr\int_{0}^{\infty}f_{0}^{2} p_{2} \,dr.
\eqno(alpha2)
$$
Alternative expressions for $\alpha_{1}$ and $\alpha_{2}$ may be obtained by
using Eqs.~\(chem1) and \(chem2) for $p_{1}(r)$ and $p_{2}(r)$, as follows:
$$
\eqalign{
       \gamma_{1}\int_{0}^{\infty}f_{0}^{2} p_{1} \,dr & =
\frac{\sigxx}{\kappa^{2}}\int_{0}^{\infty}
     \frac{d}{dr}\frac{1}{r}\frac{d(rp_{1})}{dr} \,dr\cr
   &= \frac{2\sigxx}{\kappa^{2}} p_{1}^{(1)}, \cr}
\eqno(p1)
$$
$$
\eqalign{
   \gamma_{1}\int_{0}^{\infty}f_{0}^{2} p_{2} \,dr & =
\frac{\sigxx}{\kappa^{2}}\int_{0}^{\infty}
        \frac{d}{dr}\frac{1}{r}\frac{d(rp_{2})}{dr} \,dr -\frac{\gi}{2}
          - \frac{1}{\kappa}\sigxy h_{0}(0)\cr
& =  \frac{2\sigxx}{\kappa^{2}} p_{2}^{(1)}
          -\frac{\gi}{2} - \frac{1}{\kappa}\sigxy(0) h_{0}(0),\cr}
\eqno(p2)
$$
where $\sigxy(0) = \sigxy[h_{0}(0)]$ is the normal state Hall conductivity
in a magnetic field equal to the field in the vortex core.
Then we have
$$
\alpha_{1}  = \gr \int_{0}^{\infty} (f_{0}')^{2} r \,dr
      + \frac{2\sigxx}{\kappa^{2}} p_{1}^{(1)}
              - \frac{\gi}{2}\int_{0}^{\infty} (f_{0}^{2})'  p_{2} r \,dr,
\eqno(alpha12)
$$
$$
\alpha_{2}  = - \frac{2\sigxx}{\kappa^{2}} p_{2}^{(1)}
        + \frac{1}{\kappa}\sigxy h_{0}(0)
       - \frac{\gi}{2}\int_{0}^{\infty} (f_{0}^{2})'p_{1} r \,dr  .
\eqno(alpha22)
$$
The last integral in Eq.~\(alpha12) is generally quite small, being
$O(\gi^{2},\gi\sigxy)$, and will be dropped from now on.
Finally, by using Eq.~\(nodim4) for the superfluid velocity,
we can also write Eq.~\(motion) in terms of the superfluid velocity at the
boundaries (where $f_{0}=1$), $\vsi = 2 \J_{t}/\kappa$, as
$$
\vsi\times\ez = \alpha_{1}\vl + \alpha_{2} \vl\times\ez.
\eqno(vs1)
$$
Eqs.~\(motion)-\(vs1) are the primary results of this paper.

To calculate the conductivities, we use
Faraday's law, $\langle\E\rangle=-\vl \times \B$, to obtain
$$
\J_{t} = \frac{\alpha_{1} \kappa}{2B} \langle \E\rangle
        + \frac{\alpha_{2} \kappa }{2B}
        \langle \E\rangle \times \ez.
\eqno(final)
$$
We therefore obtain the longitudinal conductivity
$$
\sigma_{xx} = \frac{\alpha_{1} \kappa}{2B},
\eqno(long)
$$
and the transverse or Hall conductivity
$$
\sigma_{xy} =\frac{\alpha_{2} \kappa}{2B}.
\eqno(hall)
$$
Returning to conventional units,  we have
$$
\sigma_{xx} = \frac{2m}{\hbar} \frac{\alpha_{1}}{8\pi \kappa^2}
             \frac{H_{c2}}{B},
\eqno(longdim)
$$
and
$$
\sigma_{xy} = \frac{2m}{\hbar} \frac{\alpha_{2} }
              {8\pi \kappa^2} \frac{H_{c2}}{B},
\eqno(halldim)
$$
with the corresponding Hall angle
$$
\tan \theta_{H} = \frac{\alpha_{2}}{\alpha_{1}}.
\eqno(angle)
$$
Therefore, we find that the Hall angle is {\it independent} of magnetic field
near $H_{c1}$.

\subhead{\bf D. Comparison to previous work}


{\it Neutral superfluids.} The dynamics of a vortex in a neutral superfluid
described by ``model A'' dynamics\refto{hohenberg77} (a nonconserved
order parameter without coupling to a conserved density) has been considered
by Onuki.\refto{onuki82} This is a limiting case of the above results,
obtained by taking $\sigxx\rightarrow \infty$, $\kappa\rightarrow\infty$,
and $\sigxy=0$;  then $p_{1}(r)=1/r$ and $p_{2}(r)=0$. Using the
expressions for $\alpha_{1}$ and $\alpha_{2}$ in Eqs.~\(alpha1) and \(alpha2),
we find
$$
\alpha_{1} = \gamma_{1} \int_{0}^{\infty}\left[ (f_{0}')^{2}
    + \frac{f_{0}^{2}}{r^{2}} \right]r\,dr,
\eqno(neutral2)
$$
$$
\alpha_{2} = -\gamma_{2}.
\eqno(neutral3)
$$
For the Galilean invariant case $\Gamma_{2}=0$; if in
addition we assume that the dissipation is small so that $\Gamma_{1}\ll1$, then
$\gamma_{1}\approx\Gamma_{1}$ and $\gamma_{2}\approx -1 $
(see Sec. II.A above). Combining these results, we arrive at the
equation of motion derived by Onuki,\refto{onuki82}
$$
(\vsi - \vl)\times\ez = \alpha_{1} \vl.
\eqno(neutral1)
$$
The left hand side is the Magnus force acting on the vortex, and the
right hand side is the viscous drag on the vortex.  If in addition
$\Gamma_{1}=0$ (so that our order parameter equation of motion corresponds
to the Gross-Pitaevskii equation), then $\alpha_{1}=0$, and the
vortex will drift with the local superfluid velocity.

An interesting feature of Eq.~\(neutral2) for $\alpha_{1}$ is that the
integral is logarithmically divergent (since $f_{0}(r)\approx 1$ for
$r\gg 1$).  According to Onuki,  the divergence should be
cut-off at a length $L$, which is either the intervortex
spacing or the wavelength of second sound.\refto{onuki82}
However, Neu\refto{neu90} has recently shown that the damping coefficient
$\alpha_{1}$ is actually velocity dependent, indicating a general breakdown
of linear response in two dimensions for the superfluid described by model A
dynamics (similar results
for the drag on a disclination line in a nematic liquid crystal have been
obtained by Ryskin and Kremenetsky\refto{ryskin91}).
The point is that for a {\it moving} vortex, there is a length scale
set by the velocity of the vortex, which in conventional units is
$L=\hbar/(2 m \gamma_{1} v_{L})$,
and it is this length which cuts off the divergence of the friction
coefficient. Therefore $\alpha_{1}\approx\gamma_{1}\ln (L/\xi)$
is perfectly well defined, in
contrast to the equilibrium energy of the vortex, which is
logarithmically divergent and will depend on the system size (or the
intervortex separation).  A more realistic model of a superfluid includes
a coupling of the order parameter to the entropy density (``model F''
dynamics\refto{hohenberg77}); as shown by Onuki,\refto{onuki83a,onuki83b} this
coupling removes the divergence in $\alpha_{1}$ by causing the dissipation
to occur in the core of the vortex.  Something similar occurs for the
charged superfluid, where there is an additional length scale set by the
conductivity of the normal fluid, as discussed by Hu and Thompson\refto{hu72};
in conventional units, this length is
$\zeta_{HT} = (4\pi\hbar\sigxx/2m\gamma_{1})^{1/2}\lambda$.  As shown in
Appendix B, if the normal fluid has a very high conductivity so that
$\zeta_{HT}\gg\xi$, then $\alpha_{1}\approx\gamma_{1}\ln(\zeta_{HT}/\xi)$;
the length $\zeta_{HT}$ cuts off the logarithmic divergence in $\alpha_{1}$.
We therefore expect linear response to hold for more realistic models
of neutral superfluids and for charged superfluids (superconductors).


{\it Paramagnetic impurities.} For a superconductor containing a
high concentration of paramagnetic
impurities, Gor'kov and \'Eliashberg\refto{gorkov68} have shown that
the parameters in the TDGL are related such that $\zeta_{HT} =\xi/\sqrt{12}$,
and  $\kappa^{-2}=48\pi\hbar \sigxx /(2m\gamma_{1})$.
Kupriyanov and Likharev\refto{kupriyanov72}
numerically integrated Eqs.~\(neworder1), \(newvectpot), and \(chem1), and
found $\alpha_{1}=0.438\gamma_{1}$.  Hu\refto{hu72b} used a hybrid method which
combined a trial order parameter solution with  numerical integration
and also found $\alpha_{1}=0.438\gamma_{1}$, correcting the earlier work of
Hu and Thompson.\refto{hu72}   The trial order parameter solution
discussed in Appendix B gives a value of $\alpha_{1}=0.436\gamma_{1}$.
Therefore, it appears that at least
in the large $\kappa$ limit the trial order parameter allows us to
calculate the transport coefficients to within 1 \%.   The results may
be extended to lower values of $\kappa$ by finding the optimal
value of $\xi_{v}(\kappa)$ using the variational principle discussed in
the Appendix.
Substituting our value of $\alpha_{1}$ into our
expression for the longitudinal conductivity, Eq.~\(longdim), we find
$$
\sigma_{xx} = 2.62\, \sigxx \frac{H_{c2}}{B},
\eqno(para1)
$$
in agreement with the results of Kupriyanov and Likharev\refto{kupriyanov72}
and Hu.\refto{hu72b}  For the Hall conductivity we obtain (in conventional
units)
$$
\sigma_{xy} = 6\, \sigxx \frac{\alpha_{2}}{\gamma_{1}} \frac{H_{c2}}{B},
\eqno(para2)
$$
where
$$
\alpha_{2} = -0.140\gamma_{2} - 0.186\left[\gamma_{2}
      + \frac{2\pi\hbar}{m}\sigxy(0)\right]
      + \frac{\pi\hbar}{m}\ln\kappa\, \sigxy(0) \frac{h_{0}(0)}{H_{c1}}.
\eqno(alpha4)
$$
There are two features of this result which are worth emphasizing.
First, the Hall conductivity contains
two contributions, one from the imaginary part of the order parameter
relaxation time $\gi$, and one from the normal state Hall conductivity.
If $\Gamma_{2}=0$ [which would produce the London acceleration equation
(see Appendix A)], then $\gi<0$, and these two contributions
have the same sign, leading to a Hall effect
in the mixed state of the same sign as in the normal state.
To get a sign change we at least require that $\gi>0$.
If, in addition, $\Gamma_{1}\ll 1$, then $\gi\approx -1$;
if we ignore the contribution from the
normal state Hall conductivity, then $\alpha_{2}=0.326$, quite different from
the value of $\alpha_{2}=1$ which we obtain for the neutral superfluid
discussed above.  Due to the screening effect of the normal fluid,
the vortex does not experience the full Magnus force as it would in a
neutral superfluid.


{\it Dirty limit.} As shown by Schmid,\refto{schmid66} in the dirty limit
the dimensionless normal state conductivity is
proportional to the real part of the order parameter relaxation time:
$\sigxx = 0.173\, \gamma_{1}$ ($\sigxx/\gamma_{1}=\Sigma$ in
Schmid's notation).  Using the results of Appendix B, we find that
$\alpha_{1}= 0.508\, \gamma_{1}$.  Therefore, in conventional units we find for
the longitudinal conductivity
$$
\sigma_{xx} = 1.47 \,  \sigxx \frac{H_{c2}}{B}.
\eqno(dirty1)
$$
Schmid obtained a similar result but with a prefactor of 1.56; we have not
been able to track down the source of this small discrepancy.
For the Hall conductivity, we have
$$
\sigma_{xy} = 2.89\, \sigxx \frac{\alpha_{2}}{\gamma_{1}} \frac{H_{c2}}{B},
\eqno(dirty2)
$$
where we again use the results of Appendix B to find
$$
\alpha_{2} = -0.178\,\gamma_{2} - 0.258\,\left[\gamma_{2}
      + \frac{2\pi\hbar}{m}\sigxy(0)\right]
      + \frac{\pi\hbar}{m}\ln\kappa\, \sigxy(0) \frac{h_{0}(0)}{H_{c1}}.
\eqno(dirty3)
$$
Fukuyama {\it et al.},\refto{fukuyama71} have derived  TDGL equations
from the microscopic BSC theory including terms which break
particle-hole symmetry, and find
$$
\gamma_{1}= {{\pi}\over {8 k_{B} T_{c}}}
                  {{\hbar^{2}}\over {2 m \xi^{2}(0)}},
\eqno(dirty4)
$$
$$
\gamma_{2}= \alpha \left({{k_{B} T_{c}}\over {\epsilon_{F}}}\right)\gamma_{1},
\eqno(dirty5)
$$
where $m$ is the effective mass of a Cooper pair in the $x-y$ plane,
$\xi(0)$ is the zero temperature correlation length in the $x-y$ plane,
$\epsilon_{F}$ is the fermi energy, and $\alpha$ is a dimensionless parameter
introduced by Fukuyama {\it et al.} which characterizes the electronic
structure of the material.  The sign of the Hall conductivity therefore
depends on the sign of $\alpha$; in this picture the sign change
would be a consequence of the detailed electronic structure of the
material.

\head{\bf IV. THERMOMAGNETIC EFFECTS IN THE LIMIT $B\ll H_{c2}$}
\taghead{4.}

\subhead{\bf A. The transport energy}

In addition to producing dissipation, moving vortices also transport
energy, in a direction which is parallel to their velocity.
This leads to thermomagnetic effects in the mixed state which
are significantly enhanced over their normal state counterparts.
In order to calculate these effects, we first note that the energy current,
which can be derived from energy conservation,\refto{schmid66}
is given by (in dimensionless units)
$$
\J^{h} = 2\E\times\h - 2\E\times \B - \frac{1}{\kappa}
     \left[ \left(\frac{\nabla}{\kappa}
        -i\A\right)\psi\left(\partial_{t}- i\Phi\right)\psi^{*}
     + \left(\frac{\nabla}{\kappa}
        +i\A\right)\psi^{*}\left(\partial_{t}+ i\Phi\right)\psi \right],
\eqno(hcurrent)
$$
where the unit of heat current is $(H_{c}^{2}/4\pi) (\hbar/2m)
(\kappa^{2}/\lambda)$.  In Eq.~\(hcurrent) we have subtracted the energy
current of the uniform background induction field, which was not
considered by Schmid.
Expressing this in terms of the gauge invariant quantities $\Q$ and
$P$, we have
$$
\J^{h} = 2( -\frac{1}{\kappa}\nabla P -
\partial_{t}\Q )\times (\nabla \times \Q)
     +\frac{2}{\kappa} \left[- \frac{1}{\kappa}(\nabla f)
        \partial_{t}f + P \Q f^{2} \right]- 2\vl B^{2}.
\eqno(hcurrent2)
$$
We again assume that the vortex translates uniformly,
and expand the order parameter and the potentials in powers of the velocity,
to obtain the local heat current
$$
\J^{h} = 2( -\frac{1}{\kappa}\nabla P +
 \vl\cdot\nabla \Q_{0})\times (\nabla \times \Q_{0})
  +\frac{2}{\kappa} \left[- \frac{1}{\kappa}(\vl\cdot\nabla f_{0})(\nabla
f_{0})
  + P \Q_{0} f_{0}^{2} \right] - 2\vl B^{2}.
\eqno(hcurrent3)
$$
Using $\nabla\times\nabla\times\Q_{0} + f_{0}^{2}\Q_{0}=0$, the first and
last terms in Eq.~\(hcurrent3) may be combined:
$$
(\nabla P)\times(\nabla\times\Q_{0}) +
   P\nabla\times\nabla\times\Q_{0}= \nabla\times(P\nabla
\times\Q_{0}),
\eqno(hcurrent4)
$$
where a vector identity has been used.  The second term on the left hand side
of Eq.~\(hcurrent3) may be written as
$$
(\vl\cdot\nabla \Q_{0})\times(\nabla\times\Q_{0})
  = \nabla\times[(\vl\cdot\Q_{0}) \nabla\times\Q_{0}] +
  \vl (\nabla\times\Q_{0})^{2} -
(\vl\cdot\Q_{0})\nabla\times\nabla\times\Q_{0},
\eqno(hcurrent5)
$$
where we have again used several vector identities.  Combining
Eqs.~\(hcurrent3)--\(hcurrent5), we have
$$\eqalign{
\J^{h} &= 2\nabla\times\left[ \left(-\frac{1}{\kappa}P + \vl\cdot\Q_{0}\right)
         \nabla\times\Q_{0}\right]\cr
    &\qquad +\frac{2}{\kappa^{2}}(\vl\cdot\nabla f_{0})(\nabla f_{0})
   +2 f_{0}^{2} (\vl\cdot\Q_{0}) \Q_{0} +  2\vl h_{0}^{2}
       - 2\vl B^{2}.\cr}
\eqno(hcurrent6)
$$
Finally, we average over all space, and note that the first term in
Eq.~\(hcurrent6) is a surface term, which vanishes.  Then we have
$$
\eqalign{
\langle \J^{h} \rangle & = \int d^{2}r \J^{h}(\r)\cr
                       & = n U_{\phi} \vl, \cr}
\eqno(hcurrent7)
$$
where $n$ is the vortex density (equal to $B/\phi_{0}$ in conventional
units and $(\kappa/2\pi) B$ in dimensionless units) and
$U_{\phi}$ is the transport energy per vortex; this combination is equal to
$$
n U_{\phi} = 2\pi \int_{0}^{\infty} [ \frac{1}{\kappa^{2}} (f_{0}')^{2}
    + f_{0}^{2} Q_{0}^{2} + 2 h_{0}^{2} ]r \,dr - 2B^{2}.
\eqno(hcurrent8)
$$
The first two terms in the integral in Eq.~\(hcurrent8) are
the kinetic energy of the superfluid
(a factor of $1/2$ comes from an angular average), while the
third term is twice the magnetic field energy. Recently,
Doria {\it et al.}\refto{doria89}  have derived a ``virial theorem'' which
shows that this combination is precisely equal to $2\H\cdot\B$.
Therefore, we find quite generally that $U_{\phi} = -(2\pi/\kappa)8\pi M$,
where $\M = (\B - \H)/4\pi$ is the spatially averaged magnetization of the
sample.  This result is true throughout the mixed state.  Near
$H_{c1}$, $B\approx 0$, so that $M\approx - H_{c1}/4\pi$.  Then
$U_{\phi}\approx (4\pi/\kappa) H_{c1}=\epsilon_{1}$, where $\epsilon_{1}$
is the line energy of the vortex.\refto{fetter}  The line energy is
calculated using a trial order parameter in Appendix B;  for
large $\kappa$, we find that the transport energy per vortex near
$H_{c1}$ is
$$\eqalign{
U_{\phi} &\approx \frac{2\pi}{\kappa^{2}}(\ln\kappa + 0.519) \cr
         &=  \left( \frac{\phi_{0}}{4\pi\lambda}\right)^{2}
            (\ln\kappa + 0.519) ,\cr}
\eqno(hcurrent10)
$$
where the units have been reinstated in the last line of Eq.~\(hcurrent10).
Therefore, the moving vortex transports an amount of energy
equal to the vortex line energy.\refto{stephen66}

Before discussing the various thermomagnetic effects, we should mention
that there have been several previous attempts to calculate the
transport energy in the low induction limit.  De Lange\refto{delange74}
and Kopnin\refto{kopnin76} both calculated the transport energy but
neglected to account for the contribution coming from the
electromagnetic field.  Hu\refto{hu76} included the contribution from
the electromagnetic field, but unfortunately never provided an explicit
derivation of his result.

\subhead{\bf B. Thermomagnetic effects}

We are now in position to calculate the thermomagnetic effects; the definitions
are summarized in Appendix C.  First, we combine Eq.~\(hcurrent7) with
Faraday's Law for the moving vortices,
$\langle\E\rangle = -\vl\times\B$, to obtain
$$
\langle \J^{h}\rangle  = (U_{\phi}/B) \langle\E\rangle\times\ez,
\eqno(hcurrent11)
$$
so that the vortex contribution to the transport coefficient
$\alpha_{xy}=U_{\phi}/B$. The normal fluid contribution to $\alpha_{xy}$
is generally several orders of magnitude smaller than the vortex contribution,
and it will therefore be omitted. Since the
energy current for a vortex is always perpendicular to the electric field, the
moving vortices do not contribute to the transport coefficient
$\alpha_{xx}$; any contribution to $\alpha_{xx}$ arises solely
from the normal fluid flow.  We may therefore write
$\alpha_{xx}=T\epsilon^{(n)}/\rho_{xx}^{(n)}$, with $\epsilon^{(n)}$ the normal
state thermopower and $\rho_{xx}^{(n)}$ the normal state resistivity.
The longitudinal thermomagnetic effects
arise primarily from the motion of the normal fluid, while the
transverse thermomagnetic effects are predominantly due to vortex motion.

We have for the Nernst coefficient $\nu$,
$$\eqalign{
\nu & = \frac{1}{TH}\rho_{xx}\alpha_{xy}\cr
    &= \frac{\phi_{0}}{TH} \frac{\hbar}{2m}
       \frac{\ln\kappa}{\alpha_{1}},\cr}
\eqno(hcurrent12)
$$
where we have used Eq.~\(longdim) for the resistivity and Eq.~\(hcurrent10)
for the transport energy.  This is the Nernst coefficient for a {\it single}
vortex; for a collection of $n=B/\phi_{0}$ noninteracting vortices,
this result should be multiplied by $n$.
The thermopower is given by
$$\eqalign{
\epsilon &=\frac{1}{T}\rho_{xx}\alpha_{xx} + H\nu\tan\theta_{H}\cr
         &= (\rho_{xx}/\rho_{xx}^{(n)})\epsilon^{(n)} + H\nu\tan\theta_{H},\cr}
\eqno(hcurrent13)
$$
with $\rho_{xx}$ the flux flow conductivity calculated above.  The vortex
motion does contribute to the thermopower if there is a nonzero Hall angle,
but under most circumstances the first term in Eq.~\(hcurrent13) is
much larger than the second term.  Therefore the thermopower in the
mixed state will generally track the behavior of the flux flow resistivity,
and not the behavior of the Hall angle.  This is in agreement with recent
measurements of the thermopower in the high temperature
superconductors;\refto{galffy90,zavaritsky91}  similar conclusions have been
reached starting from a phenomenological model.\refto{logvenov91}

\head{\bf V. VORTEX BENDING AND FLUCTUATIONS}
\taghead{5.}

So far we have limited our discussion to the motion of rectilinear vortices,
without thermal fluctuations.   However, it is straightforward to generalize
the technique discussed in Sec. III above to situations in which the vortices
are bent along the $z$-direction.  This
has been carried out by Gor'kov and Kopnin\refto{gorkov71,gorkov76}; the
addition of a complex relaxation time does not change their derivation,
so we will only outline the derivation here and refer the reader
to the original literature for the details.  First, since we have in mind
the problem of vortex motion in high temperature superconductors, we want
to allow for an anisotropic effective mass in the Ginzburg-Landau
Hamiltonian; the effective mass is $m$ in the $x-y$ plane, and $m_{z}$ along
the $z$--direction.  In this notation, $\xi$, $\lambda$, and
$\kappa=\lambda/\xi$ will denote the correlation length, penetration depth,
and Ginzburg-Landau parameter in the $x-y$ plane.
Next, we label the vortex position
by $\rl (z,t)$; $\r$ will be a position vector in the $x-y$ plane.
We assume that both the order parameter and the vector potential, which
are functions of the position $(\r,z)$ and time $t$, are functions only of the
distance away from the vortex at time $t$; e.g., $f(\r,z,t)= f(\r-\rl)$.
We again expand the order parameter, vector potential, and scalar potential
in powers of the velocity $\partial_{t}\rl$ and the
curvature $\partial_{z}\rl$ of the vortex; substituting these expansions
into the TDGL equations, we find that the zeroth order terms produce the
equilibrium Ginzburg-Landau equations, while the
first order terms produce a
set of linear inhomogeneous differential equations.  Utilizing the
translational
invariance of the equilibrium equations, we derive a solvability condition.
After evaluating this solvability condition, we arrive at the following
equation of motion for the vortex (in conventional units):
$$
 \eta_{1}\partial_{t}\rl + \eta_{2}(\partial_{t}\rl)\times\ez
   = \phi_{0} \J_{t}\times\ez
     + \eptil_{1}\frac{\partial^{2} \rl}{\partial z^{2}},
\eqno(bent)
$$
with
$$
\eta_{1} = \frac{2m}{\hbar}\left(\frac{\phi_{0}}{4\pi\lambda}\right)^{2}
                  \alpha_{1}, \qquad
\eta_{2} = \frac{2m}{\hbar}\left(\frac{\phi_{0}}{4\pi\lambda}\right)^{2}
                  \alpha_{2},
\eqno(eta)
$$
and where $\eptil_{1}=m\epsilon_{1}/m_{z}$, with $\epsilon_{1}$
the line energy of the vortex (see Appendix B).\refto{brandt77}

This equation of motion can be used to study the propagation
of helicon waves, which are elliptically polarized waves which
propagate along the $z$-direction.
Setting $\J_{t}=0$, and searching for solutions of the form
$$
\rl (z,t) = {\bf u}_{0} e^{i(kz-\omega t)},
\eqno(wave)
$$
we find (in conventional units)
$$
\omega_{\pm}= \frac{\pm\alpha_{2} -i\alpha_{1}}{\alpha_{1}^{2}+\alpha_{2}^{2}}
              \left(\frac{\hbar}{2m_{z}} \ln\kappa\right) k^{2}.
\eqno(helicon)
$$
If we set $\alpha_{1}=0$ and $\alpha_{2}=1$, then we obtain the well known
dispersion relation for helicon waves in an ideal incompressible
fluid.\refto{fetter}
However, under most circumstances $\alpha_{2}\ll\alpha_{1}$, so these waves are
overdamped, and therefore rather difficult to observe.

It is possible to include the effects of thermal fluctuations by
appealing to the fluctuation-dissipation theorem.  To the right hand side
of Eq.~\(bent) we add a fluctuating force ${\bf \zeta}(z,t)$, which
is chosen in such a manner so as to guarantee that the correct equilibrium
correlations are obtained. The resulting equation of motion
(in conventional units) is
$$
 \eta_{1}\partial_{t}\rl + \eta_{2}(\partial_{t}\rl)\times\ez
   = -\frac{\delta H_{L}}{\delta \rl} + {\bf \zeta},
\eqno(langevin)
$$
where the ``Hamiltonian'' for the vortex  line is
$$
H_{L} = \int dz \left[ \frac{\eptil_{1}}{2}\left( \frac{\partial \rl}
{\partial z} \right)^{2} -\phi_{0} (\J_{t}\times\ez)\cdot\rl \right],
\eqno(vortexham)
$$
and where the noise term has the correlations
$$
\langle  \zeta_{i}(z,t) \rangle = 0,\qquad
  \langle \zeta_{i}(z,t)\zeta_{j}(z',t')\rangle = 2\eta_{1} k_{B} T
                    \delta_{ij} \delta(z-z')\delta(t-t'),
\eqno(noise)
$$
with the brackets denoting an average with respect to the noise distribution.
We see that the first term in the Hamiltonian is the bending energy of the
vortex while the second term is essentially ``$\J_{t}\cdot \A$''; i.e.,
the interaction energy between the magnetic field and the transport current.
Similar Langevin equations for vortices in superfluid HeII have been derived
starting from model F dynamics by Onuki,\refto{onuki83a,onuki83b}
Kawasaki,\refto{kawasaki83} and Ohta, Ohta, and Kawasaki.\refto{ohta84}
Also, Ambegaokar {\it et al.}\refto{ambegaokar80} have used Langevin
equations for point vortices in two dimensions to study vortex dynamics
near the Kosterlitz-Thouless vortex unbinding transition.

We can use the Langevin equation to determine the
distance that a single
vortex line $\rl(z,t)$ wanders perpendicular to the $z$-axis in a time $t$,
in the absence of a transport current.  For simplicity we will assume that
$\eta_{2}=0$, so that the motion is purely diffusive.  Clearly the center
of mass does not move, i.e., $\langle \rl (z,t) \rangle =0$.
For the mean square displacement we have
$$
\eqalign{
\langle |\rl (z,t) -\rl (0,0)|^{2}\rangle & = 8\eta_{1} k_{B} T
    \int_{-\infty}^{\infty} \frac{d k}{2\pi} \int_{-\infty}^{\infty}
    \frac{d\omega}{2\pi} \frac{1 - e^{i(kz-\omega t)}}{(\eptil_{1} k^{2})^{2}
      + (\eta_{1} \omega)^{2}}\cr
      &= \frac{2 k_{B} T |z|}{\eptil_{1}} f(z^{2}/4D|t|),\cr}
\eqno(rms)
$$
where $D=\eptil_{1}/\eta_{1}=(\hbar/m_{z})(\ln\kappa/2\alpha_{1})$
is the diffusion constant for the vortex motion, and
where the scaling function $f(x)$ is given by
$$
 f(x)  = \frac{1}{\sqrt{\pi x}} e^{-x} + \frac{1}{\pi}
          \int_{-\infty}^{\infty} e^{-y^{2}/4x} \sin y \frac{dy}{y}.
\eqno(rms2)
$$
The last integral in Eq.~\(rms2) can be expressed in terms
of generalized hypergeometric functions, but this is not
particularly useful for our purposes.  We are primarily
concerned with the limiting behavior of the scaling function.
For $x\rightarrow 0$, the $\sin y$ may be expanded in a power series
and we find
$$
f(x) \sim \frac{1}{\sqrt{\pi x}}\left[1 + x + O(x^{2}) \right],
\eqno(rms3)
$$
while for $x\rightarrow \infty$ the integral may be calculated by steepest
descents with the result
$$
f(x) \sim 1 + \frac{2}{\sqrt{\pi x}} e^{-x} \left[ 1+ O(x^{-1})\right].
\eqno(rms4)
$$
We see that at equal times ($z\rightarrow \infty$) the mean square
displacement as a function of $z$ scales as $|z|^{1/2}$,
so that the vortex ``diffuses'' along the $z$-direction, as was first
discussed by Nelson.\refto{nelson88,nelson89}  On the other hand, if we
focus on the fixed value of $z=0$,
then at we see that the mean square displacement scales with time as
$$
\langle |\rl (0,t) -\rl (0,0)|^{2}\rangle = \frac{2 k_{B} T}{\eptil_{1}}
              \left(\frac{4 D |t|}{\pi}\right)^{1/2}.
\eqno(rms4)
$$
This is quite different from the result we would obtain
for point vortices diffusing in two dimensions, where the mean square
displacement would scale as $|t|$.   The difference is
due to the restraining effect of the line tension of the vortex.

We should stress that our vortex equation of motion was derived in the
absence of vortex pinning.  Pinning may be included in a phenomenological
fashion by including a pinning potential $V_{p}(\rl,z)$ in the Hamiltonian
for the vortex line:
$$
H_{L} = \int dz \left[\frac{\eptil_{1}}{2}\left(\frac{\partial \rl}{\partial z}
        \right)^{2} + V_{p}(\rl,z) -\phi_{0}(\J_{t}\times\ez)\cdot\rl\right].
\eqno(vortexham2)
$$
The equation of motion
would still be given by Eq.~\(langevin).  Numerical studies of
this equation have recently been carried out by Enomoto and
collaborators.\refto{enomoto92}

\head{\bf VI. DISCUSSION AND SUMMARY}
\taghead{6.}

In this paper we have derived an equation of motion for a single vortex
in a type-II superconductor in the large-$\kappa$ limit, starting from
a set of generalized TDGL equations.  This in turn allowed us to calculate
the Hall conductivity for a single vortex moving in response to
an applied transport current. There are two important features of the
results which are worth emphasizing.  First,  there are two contributions
to the Hall conductivity $\sigma_{xy}$ (and therefore to the Hall angle
$\theta_{H}$), one from the imaginary part of the order parameter relaxation
time $\gamma_{2}$, and the other from the Hall conductivity of the normal fluid
in the core of the vortex, $\sigxy (0)$.  This is different from both the
Bardeen-Stephen (BS) and Nozi\`{e}res-Vinen (NV) models, in which $\theta_{H}$
is determined entirely by the normal state Hall conductivity.
Second, $\theta_{H}$ is independent of the magnetic field.
In this regard our result resembles the behavior of $\theta_{H}$ obtained
in the NV model, but with a magnitude which depends on details such as the
normal state conductivity, the order parameter relaxation time, and so on.
This is quite different from the predictions of the BS model, in which
the Hall angle is linear in magnetic field.
Jing and Ong\refto{jing90} have recently measured the flux-flow Hall
conductivity in ${\rm NbSe}_{2}$, a material which can be prepared with
comparatively few macroscopic inhomogeneities which would serve to
pin vortices.  They find that $\theta_{H}$ in the vortex state
is field independent, in agreement with the NV model;
however, their results would also be consistent with the conclusions
of this paper.  On the other hand, Hagen {\it et al.}\refto{hagen91}
have measured the flux-flow Hall effect in thin films of the high
$T_{c}$ superconductor ${\rm Tl}_{2}{\rm Ba}_{2}{\rm CaCu}_{2}{\rm O}_{8}$
(which should also have relatively weak pinning due to enhanced
thermal fluctuations)
and find that $\theta_{H}$ has a field independent component with a
complicated temperature dependence and a component which is linear in the
magnetic field and which resembles the normal state Hall angle,
in apparent contradiction with our results.  However, it is important to
bear in mind that our derivation was for a single vortex; i.e., for
$H$ close to $H_{c1}$.  It is possible that at higher magnetic inductions
the contribution to $\sigma_{xy}$ from the Hall conductivity
of the normal fluid in the vortex core will become magnetic field dependent
(as the magnetic field in the core would then be a function of $H$).
When this is added to the field independent contribution from $\gamma_{2}$,
$\theta_{H}$ would have exactly the form suggested by the experiments of
Hagen {\it et al.}  The complicated temperature dependence of the field
independent term would be encapsulated in $\gamma_{2}$.

As discussed in the Introduction,
this work was motivated by a number of experimental observations of
an anomalous sign change in the Hall conductivity in several of the
high temperature superconductors. With results in hand, it is now time to
see whether our calculations shed any light on these puzzling
observations.  First, as previously noted, the Hall conductivity
of the vortex state will have a sign opposite to that in the normal
state if $\alpha_{2}<0$.  From the explicit calculations of $\alpha_{2}$ in
Appendix B,
we see from Eq.~(B.15) that if $\gamma_{2}>0$, there is at least the
possibility that $\alpha_{2}<0$, whereas if $\gamma_{2}<0$, it appears
that $\alpha_{2}$ is always positive.   Therefore the issue of the
sign change of the Hall conductivity hinges on whether $\gamma_{2}$ is
positive or negative.    If we choose $\gamma_{2}$ so that we generate the
London acceleration equation in the hydrodynamic limit (ensuring Galilean
invariance), then we must take $\gamma_{2}=-1$, and the Hall effect does
not change sugn.  This is in accord with the simple picture that the
vortex motion which results from the Magnus force produces a Hall conductivity
which is in the same direction as the normal state Hall effect.
If, however, we imagine deriving the TDGL equations from the microscopic
theory, along the lines of the work by Fukuyama {\it et al.},\refto{fukuyama71}
then the order parameter relaxation time depends upon the detailed
electronic structure of the material, and it is quite possible that
$\gamma_{2}>0$, producing a sign change in the Hall conductivity.

There has been a recent suggestion by Wang and Ting\refto{wang91} that
pinning forces acting on a vortex may produce backflow currents which
act in such a way so as to change the sign of the Hall angle.
It appears difficult to incorporate this effect into the present calculation.
However, there are several {\it a priori} objections to this mechanism
which are worth mentioning. First, the  observed
sign change occurs at relatively high temperatures (close to $T_{c2}$);
one might expect that at these temperatures thermal fluctuations
would tend to overwhelm the pinning forces, rendering them ineffective.
In fact, experimental studies of the Ettingshausen effect\refto{palstra90}
and the Nernst effect\refto{hagen290} in YBCO near $T_{c2}$
appear to indicate extensive flux-flow, consistent with the idea that
pinning is insignificant in this regime.
Second, from their model Wang and Ting predict that as the temperature is
lowered the pinning should cause
the longitudinal and Hall resistivities to vanish at different temperatures,
for
a fixed value of the magnetic field.  However, the recent measurements of
Luo {\it et al.}\refto{luo92} indicate that these resistivities vanish at
the same temperature.  Third, Wang and Ting begin with a hydrodynamic model
for the superfluid velocity (with the attendant shortcomings),
and incorporate the effect of pinning by including a pinning force which
depends on the fluid velocity in the normal core.   This is rather
peculiar, in that one expects the pinning force to be position dependent,
but not velocity dependent.
A crucial test of the Wang and Ting theory would be to see if a Hall
resistivity which is initially positive in the mixed state of some
material can be induced to change sign as pinning sites are artificially
introduced (by ion bombardment, for example).

Are there other measurements  which might be useful in sorting out the
sign change problem?  One might hope that measurements of the thermopower
would be useful, as the thermopower owes its existence to
particle-hole asymmetry.  Unfortunately, the thermopower in the mixed
state is dominated by the normal state
contribution, and is therefore proportional to the flux-flow resistivity
(see the discussion in Sec. IV.B and Eq.~\(hcurrent13)).
Therefore the thermopower provides very little additional information
which would be useful in piecing together the puzzle.
The observation of helicon waves in the vortex state would also be interesting;
a change in the sign of the Hall effect would cause the polarization of
the helicon waves to change direction.  However, given that these waves
are heavily overdamped, the prospects for observing this effect appear dim.
If the sign change is indeed a consequence of the electronic structure
of the material, then a more sophisticated theory should be able to
predict the existence of the sign change based on, say, band structure
calculations.  There is clearly a need for greater theoretical understanding
of the interplay between materials properties and vortex motion in
superconductors.

\head{ACKNOWLEDGMENTS}

I would like to thank S. Ullah and M. P. A. Fisher for helpful discussions.
This work was supported by NSF Grant No. DMR 89-14051, and by a fellowship
from the Alfred P. Sloan Foundation.

\head{APPENDIX A: LONDON ACCELERATION EQUATION}
\taghead{A.}

In this appendix we will derive the London acceleration equation for a
charged superfluid starting from the order parameter equation
of motion, Eq.~\(ham4).
In the London approximation,
we assume that the superfluid density is constant throughout the
fluid, and write the order parameter as
$\psi(\r,t) = n_{s}^{1/2} \exp[i\chi(\r,t)]$, with $n_{s}= |a(T)|/b$.
Substituting this into Eq.~\(ham4), and writing for the relaxation rate
$\Gamma = \Gamma_{1} + i\Gamma_{2}$, with $\Gamma_{1}$ and $\Gamma_{2}$
both real, we have for the imaginary part
$$
\hbar\partial_{t}\chi   + \mu + e^{*}\Phi + (1+\Gamma_{2})\frac{m}{2} v_{s}^{2}
         = \frac{\hbar\Gamma_{1}}{2} \nabla\cdot\vs,
\eqno(london1)
$$
where the superfluid velocity is
$$
\vs = \frac{\hbar}{m} (\nabla\chi - \frac{e^{*}}{\hbar}\A).
\eqno(london2)
$$
Taking the gradient of Eq.~\(london1), recalling that
$\E = -\nabla\Phi -\partial_{t}\A$, and defining a viscosity coefficient
$\zeta_{3}= \hbar\Gamma_{1}/2m$, we have
$$
\partial_{t}\vs + (1+\Gamma_{2})\nabla(\half v_{s}^{2}) = -\frac{1}{m}\nabla\mu
           + \frac{e^{*}}{m}\E + \zeta_{3}\nabla(\nabla\cdot\vs).
\eqno(london3)
$$

If $\Gamma_{2}=0$, then we can put this into a more familiar form by
noting that $\nabla\times\vs = -(e^{*}/m)\B$, so that
$$
\eqalign{
\nabla(\half v_{s}^{2}) &= (\vs\cdot\nabla)\vs + \vs\times(\nabla\times\vs)\cr
        &= (\vs\cdot\nabla)\vs -\frac{e^{*}}{m}\vs\times\B.\cr}
\eqno(london4)
$$
Substituting this into Eq.~\(london3), we obtain the final form of the
London acceleration equation, with a dissipative term\refto{putterman}:
$$
\partial_{t}\vs + (\vs\cdot\nabla)\vs= -\frac{1}{m}\nabla\mu
      + \frac{e^{*}}{m}(\E + \vs\times\B) + \zeta_{3}\nabla(\nabla\cdot\vs),
\eqno(london5)
$$
with the supercurrent being given by $\J_{s}= e^{*}n_{s}\vs$.

If $\Gamma_{2}\neq 0$, then Eq.~\(london3) has no simple hydrodynamic
interpretation. In fact, Eq.~\(london3) is similar to the equation
used by Vinen and Warren in their study of vortex motion in dirty
materials.\refto{vinen2}

\head{APPENDIX B: VARIATIONAL CALCULATION OF THE CONSTANTS}
\taghead{B.}

In this Appendix we will calculate the various constants which
appear in the transport coefficients,  which involve the solution
of Eqs.~\(neworder1), \(newvectpot), \(chem1) and \(chem2).  There are no known
exact solutions
to this set of equations, so generally they must be solved numerically.
However, approximate closed form solutions may be obtained  by using a
trial form for the amplitude of the order parameter $f_{0}(r)$;
since the equations
for the vector potential and the chemical potential are linear, a
sufficiently clever choice  for $f_{0}(r)$ will allow the remaining two
equations to be solved exactly.  This is the method originally due to
Schmid\refto{schmid66}, who assumed  an approximate order parameter
profile (in dimensionless units) of the form
$$
f_{0}(r) = \frac{\kappa r}{[(\kappa r)^{2} + \xi_{v}^{2}]^{1/2}},
\eqno(neworder2)
$$
where $\xi_{v}$ is a parameter which measures the healing length of the
order parameter and is numerically close to one.  An optimal choice
for $\xi_{v}$ is that which minimizes  the free energy; the dependence of
$\xi_{v}$ on $\kappa$ is then\refto{clem75}
$$
1 = \frac{\sqrt{2}}{\xi_{v}}
    [1 - K_{0}^{2}(\xi_{v}/\kappa)/K_{1}^{2}(\xi_{v}/\kappa) ]^{1/2},
\eqno(var)
$$
with $K_{0}(z)$ and $K_{1}(z)$ the standard modified Bessel functions.
In the limit of large $\kappa$, this reduces to $\xi_{v}=\sqrt{2}$, which is
the value used by Schmid,\refto{schmid66} whereas $\xi_{v}=0.935$
when $\kappa=1/\sqrt{2}$.\refto{clem75}   Upon substituting Eq.~\(neworder2)
into Eqs.~\(newvectpot) for the vector potential and Eq.~\(chem1) for the
chemical potential,
we obtain the following analytic solutions:\refto{schmid66,hu72,clem75}
$$
p_{1}(r) = \frac{R}{\xi_{v} r} \frac{K_{1}(R /\zeta)}
             {K_{1}(\xi_{v}/\zeta)},
\eqno(mu)
$$
$$
Q_{0}(r)  = - \frac{R}{\xi_{v} \kappa r}
           \frac{K_{1}(R /\kappa)}{K_{1}(\xi_{v}/\kappa)},
\eqno(Q)
$$
with the local magnetic field
$$\eqalign{
h_{0}(r) & = \frac{1}{r}\frac{\partial (r Q_{0}(r))}{\partial r}\cr
 &= \frac{1}{\xi_{v}} \frac{K_{0}(R/\kappa)}{K_{1}(\xi_{v}/\kappa)},\cr}
\eqno(b)
$$
where we have defined $R\equiv[ (\kappa r)^{2} + \xi_{v}^{2}]^{1/2}$
and $\zeta=(\sigxx/\gamma_{1})^{1/2}$.\refto{huzeta}
These quantities have the limiting behaviors
$$
p_{1}(r) =\cases{\frac{1}{r} - \frac{\kappa^{2}}{2\xi_{v}\zeta}
         \frac{K_{0}(\xi_{v}/\zeta)}{K_{1}(\xi_{v}/\zeta)} r
       + O(r^{3}),&for  $r\ll \xi_{v}/\kappa$;\cr
    \frac{\kappa}{\xi_{v}}\frac{K_{1}(\kappa r/\zeta)}{K_{1}(\xi_{v}/\zeta)},
            &for $r\gg \xi_{v}/\kappa$;\cr}
\eqno(p1origin)
$$
$$
h_{0}(r)= \cases{ h_{0}(0) - \frac{\kappa}{2\xi_{v}^{2}} r^{2} + O(r^{4}),
    &for  $r\ll \xi_{v}/\kappa$;\cr
     \frac{1}{\xi_{v}} \frac{K_{0}(r)}{K_{1}(\xi_{v}/\kappa)},
     &for $r\gg \xi_{v}/\kappa$,\cr}
\eqno(b0origin)
$$
where $h_{0}(0)$ is the field at the center of the vortex,
$$
\eqalign{
  h_{0}(0) & = \frac{1}{\xi_{v}}\frac{K_{0}(\xi_{v}/\kappa)}
               {K_{1}(\xi_{v}/\kappa)}\cr
            &=\frac{1}{\kappa}(\ln \kappa - 0.231),\cr}
\eqno(b0)
$$
and where the last line is correct in the large-$\kappa$ limit.\refto{clem75}
The core field $h_{0}(0)\approx 2H_{c1}$ in the large-$\kappa$ limit.
It is also easily verified that
$$
\eqalign{
B  &= 2\pi\int_{0}^{\infty} h_{0}(r) r \,dr \cr
                         &= \frac{2\pi}{\kappa},\cr}
\eqno(aveb)
$$
as required by flux quantization.

The equation for $p_{2}(r)$, Eq.~\(chem2),  does not appear to have an
analytic solution. However, as we are primarily interested in the
$r\rightarrow 0$
behavior, we seek an approximate solution as
follows.\refto{onuki} First, define a new function $h_{2}(r) = p_{2}(r)
+ A/r$, where $A$ is a constant to be determined.  Substituting into
Eq.~\(chem2), we have
$$
\frac{\sigxx}{\kappa^2}\frac{d}{dr}\frac{1}{r}\frac{d(rh_{2})}{dr}
 -  \gr f_{0}^{2} h_{2} = \gi f_{0}\frac{d f_{0}}{d r}
    - \frac{\sigxy}{\kappa} \frac{d h_{0}}{d r}
 - A\gr \frac{f_{0}^{2}}{r}.
\eqno(chem3)
$$
Now all of the terms on the right hand side of this equation are proportional
to $r$ as
$r\rightarrow 0$.  Therefore, we choose $A$ to eliminate these  linear terms,
so that the remaining terms are $O(r^{3})$ as $r\rightarrow0$.
Using the small $r$ behavior of $f_{0}$ and $h_{0}$, we find that we must
choose
$$
A= \frac{1}{\gr}\left[\gi + \frac{1}{\kappa^{2}} \sigxy(0)\right].
\eqno(A)
$$
With this choice of $A$, Eq.~\(chem3) is approximately homogeneous for
small $r$.  The solution is therefore
$$
h_{2}(r) =  C \frac{R}{\xi_{v} r} \frac{K_{1}(R /\zeta)}
                 {K_{1}(\xi_{v}/\zeta)},
\eqno(h2)
$$
where $C$ is a constant which must be determined from the boundary conditions.
In order that $p_{2}(0)=0$, we must have $C=A$; our approximate solution
for small $r$ is therefore
$$
\eqalign{
p_{2}(r) & =  A\left[ \frac{R}{\xi_{v} r}
  \frac{K_{1}(R /\zeta)} {K_{1}(\xi_{v}/\zeta)}
  - \frac{1}{r} \right]  \cr
         &\approx  - \frac{\kappa^{2}}{2\gr\xi_{v}\zeta}
      \frac{K_{0}(\xi_{v}/\zeta)}{K_{1}(\xi_{v}/\zeta)}
     \left[\gi + \frac{1}{\kappa^{2}} \sigxy(0)\right] r .\cr}
\eqno(chem4)
$$
We expect that this expansion will capture the small $r$ behavior in the
limit that $\zeta\rightarrow 0$.

We are now in a position to calculate the coefficients $\alpha_{1}$ and
$\alpha_{2}$, which are defined in Eqs.~\(alpha1) and  \(alpha2).
Performing the integrals, for $\alpha_{1}$ we obtain
$$
\alpha_{1} = \frac{\gr}{4} + \frac{\gr\zeta}{\xi_{v}}
   \frac{K_{0}(\xi_{v}/\zeta)}{K_{1}(\xi_{v}/\zeta)},
\eqno(newalpha1)
$$
which has the limiting behavior
$$
\alpha_{1} = \cases{\gamma_{1}\left[ \frac{1}{4} + \frac{\zeta}{\xi_{v}}
            -\frac{\zeta^{2}}{2\xi_{v}^{2}} + O(\zeta^{3})\right],
              &for  $\zeta\ll \xi_{v}$,\cr
           \gamma_{1}\left[\ln(\zeta/\xi_{v}) + 0.365  + O(\zeta^{-2})\right],
           &for $\zeta\gg \xi_{v}$.\cr}
\eqno(alpha3)
$$
For $\alpha_{2}$ we have
$$
\alpha_{2} = -\frac{\gi}{2} I(\xi_{v}/\zeta)
          -\frac{\zeta}{\xi_{v}}
   \frac{K_{0}(\xi_{v}/\zeta)}{K_{1}(\xi_{v}/\zeta)}
             \left[\gi + \frac{1}{\kappa^{2}}\sigxy(0)\right]
            + \frac{1}{\kappa} \sigxy(0) h_{0}(0),
\eqno(newalpha2)
$$
where the integral $I(z)$ is given by
$$
I(z) = \frac{2z}{K_{1}(z)} \int_{z}^{\infty} \frac{K_{1}(x)}{x^{2}}\,dx.
\eqno(iz)
$$
This integral has the limiting behavior
$$
I(z)=\cases{1 - [\frac{1}{2}(\ln z)^{2} - (\ln (2) -\gamma)\ln z] z^{2}
           + O(z^{4} (\ln z)^{2}), &for $z\ll 1$,\cr
     \frac{ 2 K_{0}(z)}{z K_{1}(z)} - \frac{4}{z^{2}}
           + \frac{16 K_{2}(z)}{z^{3} K_{1}(z)}, &for $z\gg 1$.\cr}
\eqno(iz2)
$$

Using the trial order parameter solution it is also possible to
calculate the line energy $\epsilon_{1}$  in the large $\kappa$
limit\refto{schmid66,clem75}:
$$
\epsilon_{1}\approx\frac{2\pi}{\kappa^{2}} (\ln\kappa + 0.519).
\eqno(line)
$$

\head{APPENDIX C. DEFINITION OF THE TRANSPORT COEFFICIENTS}
\taghead{C.}

In this Appendix we summarize the definitions of the transport coefficients,
for completeness.  A full discussion may be found in in Ref.~(\cite{callen}).
In the presence of an electric field ${\bf E}$ and a temperature gradient
$\nabla T$, the electrical current ${\bf J}$ and the heat current
${\bf J}^{h}$ are written as
$$\eqalign{
J_{x}  & = \sigma_{xx} E_{x} + \sigma_{xy} E_{y} + \frac{\alpha_{xx}}{T}
          \frac{\partial T}{\partial x} + \frac{\alpha_{xy}}{T}
          \frac{\partial T}{\partial y},\cr
J_{y}  & = -\sigma_{xy} E_{x} + \sigma_{xx} E_{y} - \frac{\alpha_{xy}}{T}
          \frac{\partial T}{\partial x} + \frac{\alpha_{xx}}{T}
          \frac{\partial T}{\partial y},\cr
J^{h}_{x} &= -\alpha_{xx} E_{x} - \alpha_{xy} E_{y} - \kappa_{xx}
             \frac{\partial T}{\partial x} -\kappa_{xy}
             \frac{\partial T}{ \partial y},\cr
J^{h}_{y} &= \alpha_{xy} E_{x} - \alpha_{xx} E_{y} + \kappa_{xy}
             \frac{\partial T}{\partial x} -\kappa_{xx}
             \frac{\partial T}{ \partial y},\cr}
\eqno(onsager)
$$
where the Onsager relations and rotational symmetry have been used to
simplify the equations.\refto{callen}

The isothermal Nernst coefficient
is
$$
\nu = E_{y}/H(\partial T/\partial x),
\eqno(nernst)
$$
under the conditions $J_{x} = J_{y} = \partial T/\partial y = 0$.  Then
by solving Eqs.~\(onsager), we find that the Nernst coefficient can be
expressed as
$$
\nu = \frac{1}{TH} [ \alpha_{xy}\rho_{xx} - \alpha_{xx}\rho_{xy}],
\eqno(nernst)
$$
where the resistivities are expressed in terms of the conductivities as
$$\eqalign{
\rho_{xx} &= \sigma_{xx}/(\sigma_{xx}^{2} + \sigma_{xy}^{2}),\cr
\rho_{xy} &=\sigma_{xy}/(\sigma_{xx}^{2} + \sigma_{xy}^{2}).\cr}
\eqno(halldef)
$$
Under most experimental conditions, the second term in Eq.~\(nernst)
is much smaller than the first, so for most purposes we have
$$
\nu \approx  \frac{1}{TH} \alpha_{xy}\rho_{xx}.
\eqno(nernst2)
$$

The Ettingshausen coefficient ${\cal E}$ as defined as
$$
{\cal E} = (\partial T/\partial y)/H J_{x},
\eqno(etting)
$$
under the conditions $J_{y}^{h} = J_{y} = \partial T/\partial x = 0$.
We then find that
$$
\kappa_{xx} {\cal E} = T \nu,
\eqno(etting2)
$$
which is a consequence of the Onsager relations.

The absolute thermopower $\epsilon$ is defined as
$$
\epsilon = -E_{x}/(\partial T/\partial x),
\eqno(thermo)
$$
under the conditions $J_{x} = J_{y} = \partial T/\partial y =0$.
We then find
$$
\epsilon = \frac{1}{T} [\rho_{xx}\alpha_{xx} + \rho_{xy}\alpha_{xy}].
\eqno(thermo2)
$$
Using Eq.~\(nernst2), the thermopower can be rewritten as
$$
\epsilon = \frac{1}{T} \rho_{xx} \alpha_{xx} + H\nu \tan\theta_{H},
\eqno(therm3)
$$
where $\tan\theta_{H} = \rho_{xy}/\rho_{xx}$ is the Hall angle.

\vfill\eject

\references


\refis{fetter} A. L. Fetter and P. C. Hohenberg, in {\it Superconductivity,
Vol. 2}, R. D. Parks (ed.) (Marcel Dekker Inc., New York 1969), Ch. 14.

\refis{hohenberg77} P. C. Hohenberg and B. I. Halperin, {\sl Rev. Mod.
         Phys.} {\bf 49}, 435 (1977).

\refis{putterman} For a discussion, see S. J. Putterman,
{\it Superfluid Hydrodynamics} (North-Holland, Amsterdam 1974), Ch. IX.

\refis{gross61} E. P. Gross, {\sl Nuovo Cimento} {\bf 20}, 454 (1961);
L. P. Pitaevskii, {\sl Sov. Phys. JETP} {\bf 13}, 451 (1961).  For a
discussion of the relation of these equations to vortex motion in
superfluids, see P. Nozi\`{e}res and D. Pines, {\it The Theory of
Quantum Liquids, Volume II: Superfluid Bose Liquids} (Addison-Wesley Publishing
Company, Redwood City CA 1990), Ch. 10.    A variant of these equations
has also been used to study vortex motion in the fractional quantum Hall
state; see M. Stone, \prb 42, 212, 1990.

\refis{khalatnikov} I. M. Khalatnikov, {\it An Introduction to the Theory of
Superfluidity} (Benjamin, New York 1965), Chapter 17.

\refis{stephen66} Earlier phenomenological models yield rather different
results.  For instance, M. J. Stephen, \prl 16, 801, 1966, argues that
the transport energy $U_{\phi} = TS_{\phi}$, where the transport
entropy $S_{\phi}=(\partial \epsilon_{1}/\partial T)$.  This has a much
different temperature dependence than the result derived here;  in fact,
the transport energy defined in this manner does not go to zero as
$T\rightarrow T_{c}$.

\refis{dorsey92} A. T. Dorsey and M. P. A. Fisher, \prl 68, 694, 1992.

\refis{bardeen65} J. Bardeen and M. J. Stephen, \pr 140, A1197, 1965.

\refis{nozieres66} P. Nozi\`eres and W. F. Vinen,
{\sl Phil. Mag.} {\bf 14}, 667 (1966).

\refis{vinen67} W. F. Vinen and A. C. Warren, {\sl Proc. Phys. Soc.}
 {\bf 91}, 409 (1967).

\refis{vinen2}  W. F. Vinen and A. C. Warren, Ref.~(\cite{vinen67}),
Sec. 3.2.

\refis{wang91} Z. D. Wang and C. S. Ting, \prl 67, 3618, 1991.

\refis{josephson65} B. D. Josephson, {\sl Phys. Lett.} {\bf 16}, 242 (1965).

\refis{abrahams66} E. Abrahams and T. Tsuneto, {\sl Phys. Rev.} {\bf 152},
416 (1966), have derived a TDGL from the microscopic Gor'kov equations which
contains a modified chemical potential of the form
$\tilde{\mu} = \mu - (\hbar^{2}/2m)[\nabla - i (e^{*}/\hbar)\A]^{2}$.
Although this term leads to the London acceleration equation for the superfluid
velocity, and hence appears to ensure Galilean invariance,  it leads to a
TDGL which does not relax to the correct equilibrium solution,
$\delta {\cal H}/\delta \psi^{*} = 0$.  This modified TDGL was subsequently
used by K. Maki, \prl 23, 1223, 1969, to calculate the Hall effect in the
vortex state.


\refis{hu72} C.-R. Hu and R. S. Thompson, \prb 6, 110, 1972.

\refis{ryskin91} G. Ryskin and M. Kremenetsky, \prl 67, 1574, 1991.

\refis{neu90} J. C. Neu, {\sl Physica} {\bf 43D}, 385 (1990).

\refis{enomoto92} Y. Enomoto, ``Computer simulations of a magnetic flux
motion in random impurities,'' preprint (1991); Y. Enomoto and R. Kato,
``Dynamics of a vortex line in type-II superconductors,'' preprint (1991).

\refis{huzeta} This is related to the length screening length defined
by Hu and Thompson in Ref.~(\cite{hu72}).  If we denote their length
by $\zeta_{HT}$, then $\zeta=\zeta_{HT}/\xi$, where $\xi$ is the
coherence length: $\xi= \hbar/(2m|a|)^{1/2}$.

\refis{hu72b} C.-R. Hu, \prb 6, 1756, 1972.

\refis{hu73} C.-R. Hu and R. S. Thompson, \prl 31, 217, 1973.

\refis{onuki83b} A. Onuki, {\sl Prog. Theor. Phys.} {\bf 70}, 57 (1983).

\refis{dedominicis78} C. De Dominicis and L. Peliti, \prb 18, 353, 1978.

\refis{ohta84} S. Ohta, T. Ohta, and K. Kawasaki, {\sl Physica} {\bf 128A},
               1 (1984).

\refis{onuki82} A. Onuki, {\sl J. Phys. C} {\bf 15}, L1089 (1982).

\refis{onuki83a} A. Onuki, {\sl J. Low Temp. Phys.} {\bf 51}, 601 (1983).

\refis{onuki} This calculation is quite similar to the calculation leading
to Eq.~(89) of A. Onuki, Ref.~(\cite{onuki83a}).

\refis{kawasaki83} K. Kawasaki, {\sl Physica} {\bf 119A}, 17 (1983).

\refis{gorkov71} L. P. Gor'kov and N. B. Kopnin, {\sl Zh. Eksp. Teor. Fiz.}
{\bf 60}, 2331 (1971) [{\sl Sov. Phys. JETP} {\bf 33}, 1251 (1971)].

\refis{gorkov76} L. P. Gor'kov and N. P. Kopnin, {\sl Sov. Phys. Usp.}
{\bf 18}, 496 (1976).

\refis{sonin81} E. B. Sonin, {\sl J. Low Temp. Phys.} {\bf 42}, 417 (1981).

\refis{sonin87} E. B. Sonin, {\sl Rev. Mod. Phys.} {\bf 59}, 87 (1987).

\refis{ullah91} S. Ullah and A. T. Dorsey, \prb  44, 262, 1991.

\refis{kopnin77} N. B. Kopnin and V. E. Kravtsov, {\sl Zh. Eksp. Teor.
Fiz.} {\bf 71}, 1644 (1976) [{\sl Sov. Phys. JETP} {\bf 44}, 861 (1977)].

\refis{galperin76} Yu. M. Gal'perin and E. B. Sonin, {\sl Fiz. Tverd.
Tela (Leningrad)}, {\bf 18}, 3034 (1976) [{\sl Sov. Phys. Solid State}
{\bf 18}, 1768 (1976)].

\refis{kupriyanov72} M. Yu. Kupriyanov and K. K. Likharev,
{\sl ZhETF Pis. Red.} {\bf 15}, 349 (1972) [{\sl Sov. Phys. JETP Lett.} {\bf
15}
247 (1972)].  Note that what we call $\alpha_{1}$ is called
$\gamma$ in their paper.

\refis{ambegaokar80} V. Ambegaokar, B. I. Halperin, D. R. Nelson,
and E. D. Siggia, \prb 21, 1806, 1980.


\refis{schmid66} A. Schmid, {\sl Phys. kondens. Materie} {\bf 5},
302 (1966).

\refis{gorkov68} L. P. Gor'kov and G. M. \'{E}liashberg, {\sl Zh. Eksp.
Teor. Fiz.} {\bf 54}, 612 (1968) [{\sl Sov. Phys. JETP} {\bf 27}, 328 (1968)].

\refis{fukuyama71} H. Fukuyama, H. Ebisawa, and T. Tsuzuki,
{\sl Prog. Theor. Phys.} {\bf 46}, 1028 (1971).

\refis{cyrot73} M. Cyrot, {\sl Rep. Prog. Phys.} {\bf 36}, 103 (1973).

\refis{kramer78} L. Kramer and R. J. Watts-Tobin, \prl 40, 1041, 1978.

\refis{watts81} R. J. Watts-Tobin, Y. Kr\"{a}henb\"{u}hl, and L. Kramer,
  {\sl J. Low Temp. Phys.} {\bf 42}, 459 (1981).

\refis{schon79} G. Sch\"{o}n and V. Ambegaokar, \prb 19, 3515, 1979.

\refis{rajaraman} This is similar to the argument which leads to
``zero-modes'' when considering the fluctuations about
static soliton solutions in nonlinear field theories.  See, for instance,
R. Rajaraman, {\it Solitons and Instantons} (Elsevier
Science Publishers, Amsterdam 1982), Sec. 5.5.


\refis{kim69} Y. B. Kim and M. J. Stephen, in {\it Superconductivity, Vol. 2},
R. D. Parks, ed. (Marcel Dekker, Inc., New York 1969).


\refis{clem75} J. R. Clem, {\sl J. Low Temp. Phys.} {\bf 18}, 427 (1975).

\refis{nelson88} D. R. Nelson, \prl 60, 1973, 1988.

\refis{nelson89} D. R. Nelson and H. S. Seung, \prb 39, 9158, 1989.






\refis{jing90} T. W. Jing and N. P. Ong, \prb 42, 10 781, 1990.

\refis{hagen91} S. J. Hagen, C. J. Lobb, R. L. Greene, and M. Eddy,
\prb 43, 6246, 1991.

\refis{hagen90} S. J. Hagen, C. J. Lobb, R. L. Greene, M. G. Forrester,
and J. H. Kang, \prb 41, 11 630, 1990.

\refis{chien91} T. R. Chien, T. W. Jing, N. P. Ong, and Z. Z. Wang,
\prl 66, 3075, 1991.

\refis{luo92} J. Luo, T. P. Orlando, J. M. Graybeal, X. D. Wu, and
R. Muenchausen, \prl 68, 690, 1992.

\refis{iye89} Y. Iye, S. Nakamura, and T. Tamegai, {\sl Physica} {\bf C 159},
616 (1989).

\refis{artemenko89} S. N. Artemenko, I. G. Gorlova, and Yu. I. Latyshev,
{\sl Phys. Lett.} {\bf A138}, 428 (1989).

\refis{noto76} K. Noto, S. Shinazawa, and Y. Muto, {\sl Solid State Comm.}
{\bf 18}, 1081 (1976).


\refis{palstra90} T. T. M. Palstra, B. Batlogg, L. F. Schneemeyer,
   and J. V. Waszczak, \prl 64, 3090, 1990.

\refis{hagen290} S. J. Hagen, C. J. Lobb, R. L. Greene, M. G. Forrester,
and J. Talvacchio, \prb 42, 6777, 1990.

\refis{zavaritsky91} N. V. Zavaritsky, A. V. Samoilov, and A. A. Yurgens,
{\sl Physica C} {\bf 180}, 417 (1991).

\refis{galffy90} M. Galffy, A. Freimuth, and U. Murek, \prb 41, 11 029, 1990.

\refis{logvenov91} G. Yu. Logvenov, V. V. Ryazanov, A. V. Ustinov,
and R. P. Huebener, {\sl Physica C} {\bf 175}, 179 (1991).

\refis{delange74} O. L. de Lange, {\sl J. Phys. F} {\bf 4}, 1222 (1974).

\refis{kopnin76} N. B. Kopnin, {\sl Zh. Eksp. Teor. Fiz.} {\bf 69}, 364 (1975)
[{\sl Sov. Phys. JETP} {\bf 42}, 186 (1976)].

\refis{hu76} C.-R. Hu, \prb 13, 4780, 1976.

\refis{brandt77} To be precise, we should actually identify $\eptil_{1}$
with the tilt modulus for a single vortex, $c_{44}(k=0)/n$; see
E. H. Brandt, {\sl J. Low Temp. Phys.} {\bf 26}, 735 (1977).
For short wavelength perturbations the momentum dependence of the
tilt modulus becomes important, as discussed by Brandt.
The extension to anisotropic superconductors is discussed by
E. H. Brandt and A. Sudb\o, {\sl Physica C} {\bf 180}, 426 (1991).

\refis{doria89} M. M. Doria, J. E. Gubernatis, and D. Rainer,
\prb 39, 9573, 1989.

\refis{callen} H. B. Callen, {\it Thermodynamics; an introduction to the
physical theories of equilibrium
thermostatics and irreversible thermodynamics} (Wiley, New York 1961),
Ch. 17.

\endreferences


\figurecaptions

FIG. 1.  Definition of the coordinate system $(r,\theta)$ and the
relationship between the uniform transport current $\J_{t}$, the
vortex velocity $\vl$, the average electric field $\E$,
the Hall angle $\theta_{H}$, and the
arbitrary translation vector $\d$.  The magnetic field is out of the page.

\endfigurecaptions

\endit